\documentclass[11pt,a4paper]{article}
\pdfoutput=1 
\usepackage{siunitx}
\usepackage{amsmath}
\usepackage{amssymb}
\usepackage{dsfont}
\usepackage{graphicx}
\usepackage{xspace}
\usepackage[multiple]{footmisc}
\usepackage{color}
\usepackage{epsfig,amsthm}
\usepackage{soul}
\usepackage[makeroom]{cancel}
\usepackage{mathrsfs}
\usepackage{epstopdf}
\usepackage[utf8]{inputenc}
\usepackage{hyperref}
\usepackage{slashed}
\usepackage{graphicx}
\usepackage{pict2e}
\usepackage{tikz}
\usepackage{cite}
\usepackage{float}
\usepackage[percent]{overpic}
\usepackage{multirow}
\usepackage{physics}
\usepackage{booktabs}
\usepackage{mwe}
\usepackage [autostyle, english = american]{csquotes}
\usepackage{bbm}
\usepackage{todonotes}
\MakeOuterQuote{"}

\def\lsim{\mathrel{\rlap{\lower4pt\hbox{\hskip1pt$\sim$}}
    \raise1pt\hbox{$<$}}}                
\def\gsim{\mathrel{\rlap{\lower4pt\hbox{\hskip1pt$\sim$}}
    \raise1pt\hbox{$>$}}}                

\newcommand{\beq}{\begin{eqnarray}}
\newcommand{\eeq}{\end{eqnarray}}
\newcommand{\ba}{\begin{eqnarray}}
\newcommand{\ea}{\end{eqnarray}}
\newcommand{\be}{\begin{equation}}
\newcommand{\ee}{\end{equation}}
\newcommand{\bpmatrix}{\begin{pmatrix}}
\newcommand{\epmatrix}{\end{pmatrix}}

\newcommand{\comment}[1]{\ignorespaces}

\newcommand{\s}{\newline \vspace*{-3.5mm}}

\begin{document}

\title{
	\vspace*{-3cm}
	\phantom{h} \hfill\mbox{\small KA-TP-07-2023}
	\vspace*{0.7cm}
\\[-1.1cm]
	\vspace{15mm}   
	\textbf{Electroweak Corrections to Higgs Boson Decays 
          in a Complex Singlet Extension of the SM and their
          Phenomenological Impact \\[4mm]}} 
\date{}
\author{
Felix Egle$^{1\,}$\footnote{E-mail: \texttt{felix.egle@kit.edu}},
Margarete M\"{u}hlleitner$^{1\,}$\footnote{E-mail:
	\texttt{margarete.muehlleitner@kit.edu}},
Rui Santos$^{2,3\,}$\footnote{E-mail:  \texttt{rasantos@fc.ul.pt}}
Jo\~ao Viana$^{2\,}$\footnote{E-mail:  \texttt{jfvvchico@hotmail.com}},
\\[9mm]
{\small\it
$^1$Institute for Theoretical Physics, Karlsruhe Institute of Technology,} \\
{\small\it Wolfgang-Gaede-Str. 1, 76131 Karlsruhe, Germany.}\\[3mm]
{\small\it
$^2$Centro de F\'{\i}sica Te\'{o}rica e Computacional,
    Faculdade de Ci\^{e}ncias,} \\
{\small \it    Universidade de Lisboa, Campo Grande, Edif\'{\i}cio C8
  1749-016 Lisboa, Portugal} \\[3mm]
{\small\it
$^3$ISEL -
 Instituto Superior de Engenharia de Lisboa,} \\
{\small \it   Instituto Polit\'ecnico de Lisboa
 1959-007 Lisboa, Portugal} \\[3mm]
}
\maketitle

%
%

\begin{abstract}
The complex singlet extension CxSM of the Standard Model (SM) is a
simple extension of the SM with two visible Higgs bosons in the
spectrum and a Dark Matter (DM)
candidate. In this paper we complete the computation of the next-to-leading (NLO)
electroweak (EW) corrections to on-shell and
non-loop-induced Higgs decays. Our calculations are
implemented in the code {\tt EWsHDECAY} which also includes the
relevant QCD corrections. Performing an extensive parameter scan
in the model and including all relevant theoretical and experimental
single- and di-Higgs as well as DM constraints, we obtain a viable
parameter sample. We find that current DM constraints are able to test
the model in DM mass regions where collider searches are not
sensitive. The relative EW corrections turn out to be large for scenarios with
relatively large couplings, threshold effects or small leading-order
(LO) widths. Otherwise, they are of typical EW size and can amount up to
about 20-25\%. The theory uncertainty derived from the change of the
renormalization scheme dependence then is of a few per cent. While the
NLO corrections applied in the constraints due to
  single- and di-Higgs searches impact the validity of
specific parameter points, the 
overall shape of the allowed parameter region is not yet sensitive to
the EW corrections. This picture will change with further increased
experimental precision in the future and necessitates precise predictions
on the theory side as presented in this paper.
\end{abstract}

\thispagestyle{empty}
\vfill
\newpage

\section{Introduction}
\label{sec:intro}

After the discovery of the Higgs boson by the Large Hadron Collider
(LHC) experiments ATLAS~\cite{Aad:2012tfa} and
CMS~\cite{Chatrchyan:2012ufa} open puzzles of the Standard Model
(SM) of particle physics are still awaiting their solution. One of the
most prominent ones is the question for the nature of Dark Matter 
(DM). While there has been no direct discovery of new physics so far
the precise investigation of the properties of the Higgs boson can
advance our knowledge on beyond-the-SM (BSM) physics. In
\cite{Egle:2022wmq} we investigated the Complex
Singlet extension of the SM (CxSM) where a complex scalar singlet
field is added to the model. Imposing a $\mathbb{Z}_2$ symmetry on one
of the additional scalar degrees of freedom leads to a DM candidate
which only couples to the Higgs boson and which can be tested at the
LHC in Higgs-to-invisible decays. We computed the next-to-leading
order (NLO) electroweak (EW) corrections to the Higgs decay into two DM particles and
investigated the impact on the interplay between the allowed parameter
space of the model and the constraints from the LHC experiments on the 
branching ratio of Higgs to invisible, which has been bounded 
to below 11\% by ATLAS~\cite{ATLAS:2019cid}. \s

In this paper we further advance our precision predictions for the
Higgs-to-invisible branching ratio and other observables by
completing the NLO EW corrections to all Higgs boson decays which are
on-shell and not loop induced. The calculated NLO
EW corrections are implemented in the extension {\tt sHDECAY}
\cite{Costa:2015llh} of the 
program {\tt HDECAY} \cite{Djouadi_1998,Djouadi_2019}, which computes 
the partial decay widths and branching ratios of the CxSM Higgs boson sector. 
This allows us to take into account
the state-of-the-art higher-order QCD corrections and combine them
consistently 
with our newly computed EW corrections so that we get the most precise
predictions for the CxSM Higgs boson decay widths and branching ratios
available at present. The program code has been made publicly
available and can be downloaded at the url: \\
\centerline{\url{https://github.com/fegle/ewshdecay/}}
\\

For the computation of the EW corrections we provide the possibility
to choose between different renormalization schemes to cancel the UV
divergences of the higher-order corrections. The application
and comparison of different renormalization schemes allows us to
estimate the uncertainty in the decay width predictions due to 
missing higher-order corrections. In order to investigate the impact
of our improved predictions we first perform a scan in the parameter
space of the model and keep only those data points that are in
accordance with all relevant theoretical and experimental
constraints. For this parameter sample we investigate the
sizes of the obtained EW corrections in the various decay widths for
different model set-ups, depending on which of the visible Higgs
bosons behaves SM-like. We analyze the impact of the renormalization
scheme choice. We investigate in detail the origin
of possibly large corrections and if and how they can be tamed. This
allows us to get a better understanding of the impact of EW corrections
and their related uncertainties. \s

We then move on to the investigation of the phenomenological impact of
our increased precision. For the obtained allowed parameter sample we
will study the change of the Higgs-to-invisible branching ratio
with respect to the given limits by the LHC searches. We will further
investigate how this relates to other DM observables like direct
detection and the relic density. We will also use our predictions to
estimate the impact of higher-order EW corrections on di-Higgs
production. The process is important for the determination of the
trilinear Higgs self-interaction and the experimental verification of
the Higgs mechanism. Our improved predictions will help us to get a better
understanding of the BSM physics landscape and possible DM
candidates. \s

Our paper is organized as follows. In section \ref{sec:model}, we give
a short introduction of the CxSM and set our notation. In
Section~\ref{sec:hocorrections}, we present the renormalization of the
CxSM and describe the computation of the NLO decay widths. We
furthermore present and discuss the calculation of the next-to-next-to-leading
order Higgs-to-Higgs decay width that becomes relevant for parameter
configurations with vanishing LO widths and hence also vanishing NLO widths. Section \ref{sec:hdecay}
describes the implementation of our corrections in {\tt
  sHDECAY}. In Sec.~\ref{sec:scan} we detail our numerical scan
together with the applied theoretical and experimental constraints 
before we move on to Sec.~\ref{sec:numerical} which is dedicated to the numerical
analysis. We first present the allowed parameter regions and
investigate the Dark Matter observables in our model. We then analyze
the obtained sizes of the EW corrections and investigate remaining
theoretical uncertainties estimated by applying different
renormalization schemes. Finally, we discuss the phenomenological
impact of the newly computed 
higher-order corrections. Our conclusions are given in
section~\ref{sec:concl}. 
%

\section{The Model}
\label{sec:model}

The CxSM is obtained by adding a complex singlet field to the SM
Higgs sector. We parametrize the scalar doublet field $\Phi$ and the
new singlet field $\mathbb{S}$ as
\begin{align}
\label{eq:scalar_vev_structure}
 \Phi= \begin{pmatrix}
G^+\\\frac{1}{\sqrt{2}} \left( v+H+iG^0 \right)
\end{pmatrix}, \:
\mathbb{S}=\frac{1}{\sqrt{2}}(v_{S}+S+i(v_{A}+A)),
\end{align}
where $H$, $S$ and $A$ denote real scalar fields and $G^0$ and $G^+$
the neutral and charged Goldstone bosons for the $Z$ and $W^+$ bosons, respectively. 
The vacuum expectation values $v$, $v_A$ and $v_S$ of
the corresponding fields can in general all be non-zero so that all
three scalar fields mix with each other. In our version of the model
we impose invariance of the potential under two separate
$\mathbb{Z}_2$ symmetries acting on $S$ and $A$, under which $S
\rightarrow -S$ and $A \rightarrow -A$, and additionally require $v_A$
to be zero, so that $A$ is stable and becomes the DM candidate. The
renormalizable potential is given by
\begin{align}\label{eq:scalar_potential}
\begin{split}
V=\frac{m^2}{2} \Phi^{\dagger}  \Phi + \frac{\lambda}{4}\left(  \Phi^{\dagger}  \Phi \right)^2 + \frac{\delta_{2}}{2} \Phi^{\dagger}  \Phi|\mathbb{S}|^2  +\frac{b_{2}}{2}|\mathbb{S}|^2+\frac{d_2}{4}|\mathbb{S}|^4 + \left( \frac{b_1}{4}\mathbb{S}^2+c.c.\right),
\end{split}
\end{align}
with all parameters chosen to be real. We choose $v_S \neq 0$ so that
the other  $\mathbb{Z}_2$ symmetry is broken and $S$ and $H$ mix with each
other.\footnote{Compared to the CxSM with a DM candidate that was
  discussed in \cite{Costa:2015llh} and implemented in the code {\tt
    sHDECAY} \cite{shdecaywebpage}, we here choose a different model by
  imposing an extra $\mathbb{Z}_2$ 
  symmetry on $S$ (which is subsequently broken). The additional term
  $a_1 \mathbb{S}$ in the potential of \cite{Costa:2015llh} does not
  appear then in our potential Eq.~(\ref{eq:scalar_potential}), {\it
    i.e.} $a_1=0$ here. \label{footnote1}} The mass eigenstates of
the CP-even field $h_i$ ($i=1,2$) are obtained through the rotation
\begin{align}
\label{eq:rotationdef}
\begin{pmatrix}
h_1 \\ h_2
\end{pmatrix}
=R_{\alpha}
\begin{pmatrix}
H \\ S \\
\end{pmatrix},
\end{align}
with 
\begin{align}
\label{eq:DefRotationmatrix}
R_\alpha=\begin{pmatrix}
\cos \alpha & \sin\alpha  \\
-\sin \alpha & \cos \alpha  \\
\end{pmatrix} \equiv
\begin{pmatrix}
c_{\alpha} & s_{\alpha} \\
-s_{\alpha} & c_{\alpha} \\
\end{pmatrix} .
\end{align}
The mass matrix $\mathcal{M}$ in the gauge basis $(H,S)$ reads
\begin{align}
\label{eq:massmatrix}
\mathcal{M}= \begin{pmatrix}
\frac{v^2\lambda}{2} & \frac{\delta_{2}vv_{S}}{2} \\
\frac{\delta_{2}vv_{S}}{2} & \frac{d_{2}v_{S}^2}{2} \\
\end{pmatrix} + \begin{pmatrix}
\frac{T_{1}}{v} & 0 \\
0 & \frac{T_{2}}{v_{S}}\\
\end{pmatrix},
\end{align}
with the tadpole parameters $T_1$ and $T_2$ defined via the
minimization conditions, 
\begin{subequations}
\label{eq:minimizationcond}
\begin{align}
\frac{\partial V}{\partial v}  \equiv T_1 \; \Rightarrow \; \frac{T_{1}}{v}&= \frac{m^2}{2}+\frac{\delta_2v_{S}^2}{4}+ \frac{v^2\lambda}{4},\\
\frac{\partial V}{\partial v_S} \equiv T_2 \; \Rightarrow \; \frac{T_{2}}{v_{S}}&= \frac{b_1+b_2}{2}+\frac{\delta_2v^2}{4}+ \frac{v_{S}^2d_2}{4} \, .
\end{align}
\end{subequations}
At tree level we have $T_i = 0$ ($i=1,2$). The mass of the DM candidate $A$ is given by
\begin{align}
m_{A}^2=\frac{-b_1+b_2}{2}+\frac{\delta_2v^2}{4}+ \frac{v_{S}^2d_2}{4}=-b_1 + \frac{T_2}{v_S},
\end{align}
and the remaining mass values are the eigenvalues of the mass matrix
${\cal M}$,
\begin{align}\label{eq:massmatrixdiag}
D_{hh}^{2} \equiv R_{\alpha} \mathcal{M} R^{T}_{\alpha} \;,\qquad
  D_{hh}^2=\mbox{diag}(m_{h_1}^2,m_{h_2}^2) \;.
\end{align}
The scalar spectrum of the CxSM consists of two visible Higgs bosons
$h_1$ and $h_2$, where by definition $m_{h_1} < m_{h_2}$, and a DM
scalar $A$. One of the $h_i$ ($i=1,2$) is the SM-like 125~GeV Higgs
boson. The mixing of the two scalars leads to a modification of their
couplings to the SM particles given by the factor $k_i$,
\begin{align}
\label{eq:HiggsVbosoncoupling}
g_{h_{i} SM \, SM}=g_{H_{\mathrm{SM}}SM \, SM} k_{i} \;, \; \;
  k_{i}\equiv \begin{cases}
\begin{array}{ll} \cos\alpha\;, & i=1 \\ -\sin\alpha \;,
                                &i=2 \end{array}\end{cases},
\end{align}
where $g_{H_{\mathrm{SM}} SM \, SM}$ denotes the SM coupling between
the SM Higgs and the SM particle $SM$. As input parameters of our
model we choose
\begin{align}
v\,,\ v_{S}\,,\ \alpha\,,\ m_{h_1}\,,\ m_{h_2}\,,\ m_{A} \,,
\end{align}
in terms of which the parameters of the potential are given by
\begin{subequations}
\label{eq:inputparameters}
\begin{align}
\lambda&= \frac{m_{h_1}^2+ m_{h_2}^2+ \cos2 \alpha (m_{h_1}^2-m_{h_2}^2)}{v^2} \\
d_2&= \frac{m_{h_1}^2+ m_{h_2}^2+ \cos2 \alpha (m_{h_2}^2-m_{h_1}^2)}{v_{S}^2} \\
\delta_2&= \frac{(m_{h_1}^2-m_{h_2}^2)\sin2 \alpha}{v v_S} \\
m^2&=\frac{1}{2}\left( \cos2\alpha (m_{h_2}^2-m_{h_1}^2) -\frac{v(m_{h_1}^2+m_{h_2}^2)+v_S (m_{h_1}^2 -m_{h_2}^2)\sin2\alpha}{v} \right) \\
b_2&=\frac{1}{2}\left( 2 m_{A}^2 - m_{h_1}^2 -m_{h_2}^2 + \cos2\alpha (m_{h_1}^2-m_{h_2}^2) -\frac{v(m_{h_1}^2-m_{h_2}^2)\sin2\alpha}{v_S} \right) \\
b_1&=-m_{A}^{2} \,.
\end{align}
\end{subequations}
Two of the input parameters are fixed; the mass of the SM-like Higgs
boson has to be equal to 125.09~GeV, and the doublet VEV $v$ is given
by $v=1/\sqrt{\sqrt{2}G_F}\approx 246.22$~GeV, where
$G_F$ denotes the Fermi constant, which we choose in the following as
input parameter instead of $v$. 

\section{Electroweak Corrections}
\label{sec:hocorrections}
In addition to the already calculated EW corrections to the Higgs
decays into two DM particles we present in this paper our new
calculation of the EW corrections to the remaining non-loop-induced
on-shell decays of the visible $h_i$. These are, if kinematically allowed, the
Higgs-to-Higgs decays $h_2 \to h_1 h_1$, the decays into massive gauge
bosons, $h_i \to VV$ ($V=W^\pm, Z$) and the decays into fermions $h_i
\to f\bar{f}$. Note that we do not include EW corrections to the
loop-induced decays into photons or gluons, as they would be of
two-loop order. We furthermore do not include EW corrections to
off-shell decays, so that for the SM Higgs we do not consider
corrections to off-shell decays into massive gauge bosons. And also
for heavier Higgs bosons below the top-pair threshold we do not
include EW corrections into off-shell tops. \s

The higher-order corrections involve UV divergences that have to be
cancelled through the process of renormalization. We replace the bare
fields and parameters of our Lagrangian by the renormalized ones and
their corresponding counterterms. For the isolation of
the divergences we work in $D=4-\epsilon$ dimensions so that the
divergences appear as poles in $\epsilon$. The finite parts of the
counterterms are determined by the chosen renormalization scheme. We
offer several schemes which fulfill the
following requirements\footnote{We follow here the same guidelines that
we applied in the computation of the EW corrections to the
2-Higgs-Doublet-Model (2HDM) \cite{Krause:2016oke,Krause:2016xku,Krause:2018wmo,Krause_2020} and the Next-to-2HDM (N2HDM)
\cite{Krause:2017mal,Krause:2019oar}. Renormalization schemes
respecting some or all of the chosen criteria have also been discussed in \cite{Denner:2016etu,Altenkamp:2017ldc,Altenkamp:2017kxk,Denner2018,Fox:2017hbw,Grimus:2018rte}.}:
\begin{itemize}
\item[-] We require on-shell (OS) renormalization conditions wherever
  possible.
\item[-] The chosen renormalization schemes preserve
gauge-parameter-independent relations between the input parameters and
the computed observables.
\item[-] If possible, renormalization schemes that lead to unnaturally
  large corrections are avoided. We call good renormalization schemes
  in this context ``numerically stable''.
\item[-] If possible, process-dependent renormalization schemes, {\it
    i.e.}~renormalization schemes that depend on a physical process,
  are avoided.
\end{itemize} 
The reason for the latter condition is the exclusion of parameter
scenarios where the chosen process is kinematically not
allowed. Numerically stable conditions reflect a good convergence of
the higher-order corrections. Gauge-parameter-independent relations
allow to relate different observables to each other. On-shell
conditions can make use of measured experimental parameters like the
masses of the particles. \s

\subsection{Renormalization of the CxSM}
In the following we present the renormalization of the CxSM. Since the
renormalization conditions have already been presented in the
literature or can be taken over from other models, we restrict us here
to the minimum and refer for further details to the literature. Our
main goal here is setting our notation for the various renormalization
schemes.

\subsubsection{The Scalar Sector}
In \cite{Egle:2022wmq} we computed the NLO EW corrections to the Higgs
boson decays into DM pairs, $h_i \to AA$ and introduced the
renormalization of the scalar sector. Here we only list the main
ingredients and refer to \cite{Egle:2022wmq} for further
details. Electroweak corrections to the scalar sector requires the
field and mass renormalization of $h_1$, $h_2$ and $A$, the
renormalization of the tadpoles $T_1$ and $T_2$, of the mixing angle
$\alpha$ and of the singlet VEV $v_S$. We apply the following
renormalization conditions:
\begin{itemize}
\item {\it Mass and field renormalization of $h_1, h_2, A$:} For this, we choose OS
conditions. 
\item {\it Tadpole renormalization:} The tadpole renormalization is
  related to the way we choose the VEVs at 1-loop order so that the
  minimum conditions hold. We follow the scheme proposed by Fleischer
  and Jegerlehner \cite{Fleischer:1980ub}  for the SM. In this way, all
  counterterms related to physical quantities will become gauge
  independent. The obtained VEV is the true VEV of the
  theory. Note that in this scheme the self-energies contain
  additional tadpole contributions, and in the virtual vertex
  corrections additional tadpole contributions have to be taken into
  account if the resulting coupling that comes along with it exists in
  the CxSM ({\it cf.}~the appendix of \cite{Krause:2016oke} for a
  discussion in the 2HDM). 
\item {\it Renormalization of the mixing angle $\alpha$:} We use the
  OS-pinched and the $p_*$-pinched scheme that was introduced in
  \cite{Krause:2016oke} for the 2HDM and in \cite{Krause:2017mal} for
  the N2HDM. It is based on the 
  scheme proposed in \cite{Pilaftsis:1997dr,Kanemura:2004mg} which
  relates the mixing angle 
  counterterm to the field renormalization matrix constant. Combined
  with the Fleischer-Jegerlehner scheme for the treatment of the
  tadpoles and the electroweak VEV and the pinch technique
  \cite{Cornwall:1989gv, Papavassiliou:1994pr} to extract 
  the gauge-independent part unambiguously, the two schemes introduce a
  gauge-parameter-independent counterterm for $\alpha$. The two schemes differ
  only in the choice of the value for the squared external momenta in
  the pinched self-energies, which is
  either the mass squared of the incoming scalar or the squared mean
  of the masses of the incoming and outgoing scalar, respectively. 
\item{\it Renormalization of the singlet VEV $v_S$:} For the
  renormalization of the singlet VEV $v_S$ we offer the choice between
  the process-dependent scheme and the zero external momentum (ZEM)
  scheme \cite{Azevedo:2021ylf}. In the process-dependent scheme, we
  choose either $h_1 \to 
  AA$ or $h_2 \to AA$ for the renormalization of $v_S$ as they contain
  in their $h_{1/2} AA$ coupling the parameter $v_S$. In order to be
  used for the
  renormalization, they have to be kinematically allowed and must be
  different from the process that we want to renormalize. The
  counterterm $\delta v_S$ of $v_s$ is then extracted from the 
  process by demanding that the NLO amplitude is equal to the
  leading-order (LO) amplitude.  We call the renormalization scheme
  based on the $h_1 \to AA$ decay 'OSproc1' and the one based on $h_2
  \to AA$ 'OSproc2'. Since it is not guaranteed
  for each valid parameter point of the model that these processes are
  kinematically allowed, we also offer the ZEM scheme introduced in
  \cite{Azevedo:2021ylf} for both process choices, called 'ZEMproc1'
  and 'ZEMproc2' in the following. It avoids the kinematic restrictions of
  the OS decays into OS final states by setting the squared
  external momenta to zero in the decay process that is used for the
  renormalization. We ensure this counterterm to be gauge independent
  by using the pinched versions of the self-energies in the wave
  function renormalization constants that occur in the ZEM
  counterterm of $v_S$. 
\end{itemize}

\subsubsection{The Gauge and the Fermion Sector \label{sec:gaugesec}}
The mass and field renormalization counterterm constants, $\delta m_V$
and $\delta Z_V$ of the massive gauge bosons $V=W^\pm, Z$ (we do not
need to renormalize 
the photon) are obtained in the OS scheme. \s

The counterterm $\delta Z_e$ for the
electric charge is determined from the photon-electron-positron
($\gamma e\bar{e}$) vertex
in the Thomson limit. In the computation of the decay widths we use
the $G_\mu$ scheme \cite{Bredenstein:2006rh} in order to improve their
perturbative behavior by absorbing a large universal part of the 
EW corrections in the LO decay width. To avoid double counting we have to subtract the corresponding NLO part from the explicit EW
corrections of our NLO 
calculation. We achieve this by redefining the charge renormalization
constant accordingly. For details, we refer to
\cite{Azevedo:2021ylf}. 
The mass and field renormalization constants $\delta m_f$ and $\delta
Z_f$ are chosen in the OS scheme. For details, see~\cite{Krause:2016oke}. \s

In Table~\ref{tab:renscheme} we summarize for convenience the various
renormalization schemes used in the computation of the EW-corrected
decay widths. 

\begin{table}[ht!]
\begin{center}
\begin{tabular}{|c|c|} \hline 
{\bf Scalar Sector} & \\ \hline
$\delta m_\phi$, $\delta Z_\phi$ ($\phi=h_1,h_2,A$) & OS \\ \hline
$\delta \alpha$ & OS-pinched \\
& $p_*$-pinched \\ \hline
$\delta v_S$ & OSproc1 (OS $h_1 \to AA$)\\
& OSproc2 (OS $h_2 \to AA$) \\
& ZEMproc1 (ZEM $h_1 \to AA$) \\ 
& ZEMproc2 (ZEM $h_2 \to AA$) \\ \hline \hline
{\bf Gauge Sector} & \\ \hline
$\delta m_V$, $\delta Z_V$ ($V=W^\pm,Z$) & OS \\ \hline
$\delta Z_e$ & $G_\mu$ scheme \\ \hline\hline
{\bf Fermion Sector} & \\ \hline
$\delta m_f$, $\delta Z_f$ & OS \\ \hline
\end{tabular}
\caption{Summary of the renormalization schemes applied in the EW
  corrections to the non-loop-induced on-shell decay widths of the
  CxSM Higgs boson decays. For details, see text and
  Refs.~\cite{Krause:2016oke,Azevedo:2021ylf,Egle:2022wmq}. \label{tab:renscheme}}
\end{center}
\end{table}

\subsection{The EW-Corrected Decay Widths at NLO}
The NLO decay width $\Gamma^{\text{NLO}}$ for the decay of the scalar
Higgs $h_i$ into two final state particles $XX$ can be written as the sum of
the LO width $\Gamma^{\text{LO}}$ and the one-loop corrected decay
width $\Gamma^{(1)}$, 
\beq
\Gamma^{\text{NLO}}_{h_i \to XX} = \Gamma^{\text{LO}}_{h_i\to XX} +
\Gamma^{(1)}_{h_i \to XX} \;.
\eeq 
The one-loop corrected $\Gamma^{(1)}$ is obtained from the
interference of the LO and the NLO amplitudes ${\cal
  M}^{\text{LO}}$ and ${\cal M}^{\text{NLO}}$, respectively, 
\beq
\Gamma^{\text{NLO}} \propto 2 \mbox{Re} \left( {\cal M}^{\text{LO}} {\cal
  M}^{\text{NLO}*} \right)
\eeq
so that 
\beq 
\Gamma^{\text{NLO}}_{h_i \to XX} = \Gamma^{\text{LO}} + \Gamma^{(1)}
\propto |{\cal M}_{h_i \to 
  XX}^{\text{LO}}|^2 +  2 \mbox{Re} \left( {\cal M}^{\text{LO}}_{h_i \to
  XX} {\cal M}^{\text{NLO}*}_{h_i \to XX} \right) \;.
\eeq
We get for the individual LO decay widths $\Gamma^{\text{LO}}_{h_i \to
  XX}$ of the $h_i$ Higgs decays into a lighter Higgs pair, two
massive gauge bosons and a fermion pair, respectively,
\beq
\Gamma^{\text{LO}}_{h_i \to \phi_a \phi_a} = \frac{G_F m_Z^4 
\tilde{g}_{h_i \phi_a \phi_a}^2}{16 \sqrt{2}\pi m_{h_i}} \sqrt{1- 4
\frac{m_{\phi_a}^2}{m_{h_i}^2}} \;,
\eeq
with $h_i=h_{1,2}$ and $\phi_a=h_1, A$,
\beq
\Gamma^{\text{LO}}_{h_i \to VV} = \frac{G_F \tilde{g}_{h_i VV}^2}{8 \sqrt{2}
  \delta_{\text{s}}\pi m_{h_i}} (m_{h_i}^4 - 4 m_{h_i}^2 m_V^2 + 12 m_V^4)
\sqrt{1-\frac{4m_V^2}{m_{h_i}^2}} \;,
\eeq
where $V=Z,W^\pm$ and $\delta_{\text{s}}=1\, (2)$ for $V=W^\pm\, (Z)$, and
\beq
\Gamma^{\text{LO}}_{h_i \to ff} = \frac{N_c G_F m_f^2 m_{h_i} \tilde{g}_{h_i
    f\bar{f}}^2}{4\sqrt{2} \pi} \left(1-\frac{4m_f^2}{m_{h_i}^2}\right)^{\frac{3}{2}} \;,
\eeq
with the color factor $N_c=3 (1)$ for quarks (leptons). The couplings are  normalized as
\begin{align}
\tilde{g}_{h_i \phi_a \phi_a}= \frac{g_{h_i  \phi_a \phi_a} v}{m_Z^2},
  \quad  \tilde{g}_{h_i VV}=\frac{g_{h_i VV} v}{2m_V^2}, \quad \tilde{g}_{h_i
    f\bar{f}}=\frac{g_{h_if\bar{f}} v}{m_f}.
\end{align}
Since we consider only on-shell decays, the corresponding mass values
must be such that $m_{h_i} \ge 2 m_X$ ($X=\phi_a,V,f$). The $h_i$
coupling factors $g_{h_i SM \, SM}$ to the SM particles $SM \equiv f,
V$ were given in Eq.~(\ref{eq:HiggsVbosoncoupling}). The trilinear 
couplings between the scalar particles that we need for the 
computation of the NLO decay widths, are given by, 
\beq
g_{h_1 h_1 h_1} &=& 3 m_{h_1}^2 \frac{v_S c_\alpha^3+v s_\alpha^3}{v
  v_S} \label{eq:h1h1h1}\\
g_{h_1 h_1 h_2} &=& \frac{(2m_{h_1}^2+ m_{h_2}^2)s_{\alpha}
  c_{\alpha}(v s_{\alpha}- v_S c_{\alpha})}{v v_S}  \label{eq:h1h1h2} \\
g_{h_1h_2 h_2}&=& \frac{(m_{h_1}^2+ 2 m_{h_2}^2)s_{\alpha}
  c_{\alpha}(v c_{\alpha} + v_S s_{\alpha})}{v v_S}  \label{eq:h1h2h2}
\\
g_{h_2 h_2 h_2} &=& 3 m_{h_2}^2 \frac{v c_\alpha^3- v_S s_\alpha^3}{v
  v_S} \label{eq:h2h2h2}\\
g_{h_i AA}&=& \frac{m_{h_i}^2}{v_s} \begin{cases}
\begin{array}{ll} s_\alpha\,, & i =1 \\ c_\alpha \,, & i=2 \end{array}
\end{cases} \label{eq:hiAA} \;.
\eeq  
The one-loop correction $\Gamma^{(1)}$ consists of the virtual
corrections, the counterterm contributions and - if applicable - the
real corrections. The counterterms cancel the UV divergences and the
real corrections the infrared (IR) divergences, if they are encountered in
the virtual corrections. This happens if a massless particle is
running in the loop. For example in Higgs decays into charged $W^\pm$
bosons a photon can be exchanged in the loop diagrams. The IR
divergences are cancelled by the real corrections, that contain
bremsstrahlung contributions where a photon is radiated from the
charged initial and final state particles, and of diagrams that
involve a four particle vertex with a photon. For details, see {\it
  e.g.}~Ref.~\cite{Krause:2016oke} which describes the procedure for
the 2HDM that can be easily translated to our model. The virtual
corrections are built up by the pure vertex corrections and the
external leg corrections. The latter vanish due to the chosen OS
renormalization of the external particles. The vertex corrections
comprise all possible 1-particle irreducible diagrams. The
counterterm contribution contains the involved wave function
renormalization constants and parameter (couplings, masses, mixing
angles) counterterms. \s

The calculations of the NLO corrections were performed by two
independent calculations, using 
 {\tt FeynArts~3.10}~\cite{Kublbeck:1990xc,Hahn:2000kx} and {\tt
   FeynCalc~9.3.1}~\cite{Mertig:1990an,Shtabovenko:2016sxi}.  
 Loop integrals were computed using {\tt LoopTools}~\cite{Hahn:1998yk, vanOldenborgh:1989wn}.
 The model file was generated independently using  {\tt
   SARAH~4.14.2}~\cite{Staub:2009bi,Staub:2010jh,Staub:2012pb,Staub:2013tta,Staub:2015kfa}
 and {\tt
   FeynRules}~\cite{Christensen:2008py,Degrande:2011ua,Alloul:2013bka}. Both
 calculations found agreement between the results.  

\subsection{EW-Corrected Decay Width at NNLO}
\begin{figure}[t!]
    \centering
    \includegraphics[width=0.8\textwidth]{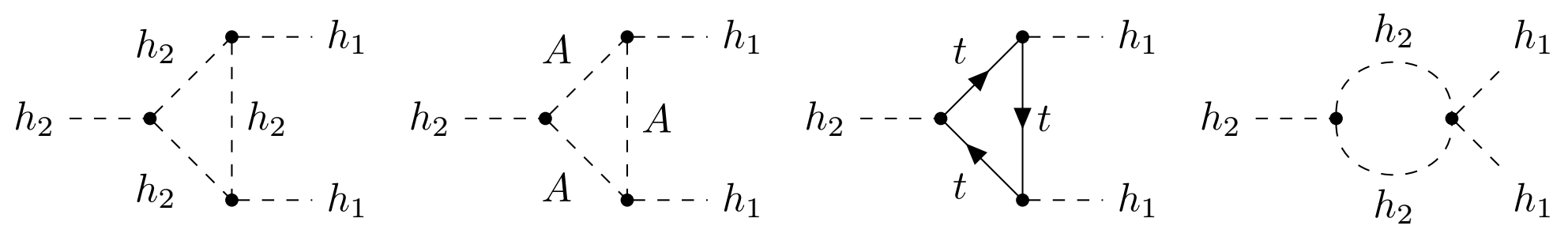}
    \caption{Sample diagrams contributing to the NLO decay width $h_2
      \to h_1 h_1$ for $\tan\alpha= v_S/v$.\label{fig:nnlodiags}}
\end{figure}
It turns out that for certain parameter configurations the LO decay
width vanishes. This can happen for the decay $h_2 \to h_1 h_1$. The
LO decay amplitude is given by
\beq
{\cal M}^{\text{LO}}_{h_2 \to h_1 h_1} = \frac{\sin 2\alpha}{2 v_S
  v} (2m_{h_1}^2 + m_{h_2}^2) (v \sin\alpha - v_S \cos \alpha) \;.
\eeq
As expected, this amplitude vanishes in the SM-like limit $\alpha =
\pi/2 (0)$ where the portal coupling $\delta_2$ vanishes, the $h_1$ ($h_2$) 
decouples and $h_2$ ($h_1$) is the SM Higgs boson. The amplitude vanishes,
however, also for parameter configurations with
\beq
\tan \alpha = \frac{v_S}{v} \;.
\eeq
Since the NLO decay width is proportional to the LO decay width it also 
vanishes. For ${\cal M}_{h_2 \to h_1 h_1}^{\text{LO}}=0$ we have
\beq
\Gamma_{h_2 \to h_1 h_1}^{\text{NLO}} \propto |{\cal M}^{\text{LO}}_{h_2
  \to h_1 h_1}|^2 + 2 \mbox{Re} \left( {\cal M}^{\text{LO}}_{h_2 \to
    h_1 h_1} {\cal M}_{h_2 \to h_1 h_1}^{\text{NLO} \, \dagger}
\right) \stackrel{{\cal M}^{\text{LO}}=0}{=} 0 \;.
\eeq
In the SM-like limit also the NLO amplitude ${\cal M}_{h_2 \to h_1
  h_1}^{\text{NLO}}$ is zero. However, for $\tan \alpha =
\frac{v_S}{v}$ we can have non-vanishing contributions to ${\cal M}_{h_2 \to h_1
  h_1}^{\text{NLO}} \ne 0$ ({\it cf.}~Fig.~\ref{fig:nnlodiags}) so
that we obtain a non-vanishing decay width which is of 
next-to-next-to-leading order (NNLO) and given by
\beq
\Gamma_{h_2 \to h_1 h_1}^{\text{NNLO}} &\propto& |{\cal M}^{\text{LO}}_{h_2
  \to h_1 h_1}|^2 +
|{\cal M}^{\text{NLO}}_{h_2 \to h_1 h_1}|^2  \nonumber \\
&+& 2 \mbox{Re} \left( {\cal M}^{\text{LO}}_{h_2 \to
    h_1 h_1} {\cal M}_{h_2 \to h_1 h_1}^{\text{NLO} \, \dagger} +
{\cal M}^{\text{LO}}_{h_2 \to
    h_1 h_1} {\cal M}_{h_2 \to h_1 h_1}^{\text{NNLO} \, \dagger} \right) \nonumber \\
&\stackrel{\tan\alpha=v_S/v}{=}& |{\cal M}^{\text{NLO}}_{h_2 \to h_1
  h_1}|^2 \;.
\label{eq:partnnlo}
\eeq
In contrast to the computation of the NLO decay width we now also have
to take into account the imaginary parts of the NLO amplitude ${\cal
  M}^{\text{NLO}}_{h_2 \to h_1 h_1}$. While the amplitude is still
UV-finite, we have to ensure that the imaginary parts do not destroy
the gauge-parameter independence of the NNLO amplitude. This is achieved by
taking into account also the imaginary part of the wave function
renormalization constant. It cancels the gauge-parameter dependence of
the imaginary part of the leg contribution that is left over after
applying our renormalization conditions. We note, that the consideration of the
imaginary part of the wave function renormalization constant to get
the NNLO amplitude gauge-parameter independent does not have any effect
on our previous calculation of the NLO widths, as here we always take
the real part of the NLO amplitude. \s

In order to calculate the full NNLO decay width when we move away from
$\tan\alpha=v_S/v$ we would also need to take into account the term $2
\mbox{Re}({\cal M}^{\text{LO}}_{h_2 \to h_1 h_1} {\cal M}_{h_2 \to h_1
  h_1}^{\text{NNLO} \, \dagger})$ in Eq.~(\ref{eq:partnnlo}). For
this, we would need to calculate the NNLO decay amplitude ${\cal M}_{h_2 \to h_1
  h_1}^{\text{NNLO}}$, which is beyond the scope of this
work. Instead, we only include the approximate NNLO width, given by
\beq
\Gamma_{h_2 \to h_1 h_1}^{\text{NNLO, \, approx}} &\propto& |{\cal
  M}^{\text{LO}}_{h_2 
  \to h_1 h_1}|^2 +
|{\cal M}^{\text{NLO}}_{h_2 \to h_1 h_1}|^2 + 
2 \mbox{Re} \left( {\cal M}^{\text{LO}}_{h_2 \to
    h_1 h_1} {\cal M}_{h_2 \to h_1 h_1}^{\text{NLO} \, \dagger} \right) \;.
\label{eq:partialnnlo}
\eeq
Sufficiently away from $\tan\alpha=v_S/v$, the NLO width should 
dominate about the NNLO one, and the incomplete NNLO calculation
should not add much to the theoretical uncertainty. Sufficiently close
to $\tan\alpha=v_S/v$ the term $2 \mbox{Re}({\cal M}^{\text{LO}}_{h_2 \to
    h_1 h_1} {\cal M}_{h_2 \to h_1 h_1}^{\text{NNLO} \, \dagger})$
  should not contribute much to the NNLO width as the LO amplitude is
  close to zero so that the approximation Eq.~(\ref{eq:partialnnlo})
  should be good enough. In the intermediate region, however, only the
  complete NNLO computation would give information on the relative
  importance of the missing NNLO contribution, $2 \mbox{Re}({\cal M}^{\text{LO}}_{h_2 \to
    h_1 h_1} {\cal M}_{h_2 \to h_1 h_1}^{\text{NNLO} \, \dagger})$, compared to the one
  taken into account by us, $|{\cal M}^{\text{NLO}}_{h_2 \to h_1
    h_1}|^2$, and would hence give information on the associated
  theoretical uncertainty due to the incomplete NNLO calculation. \s

\begin{figure}[t!]
    \centering
    \includegraphics[width=0.4\textwidth]{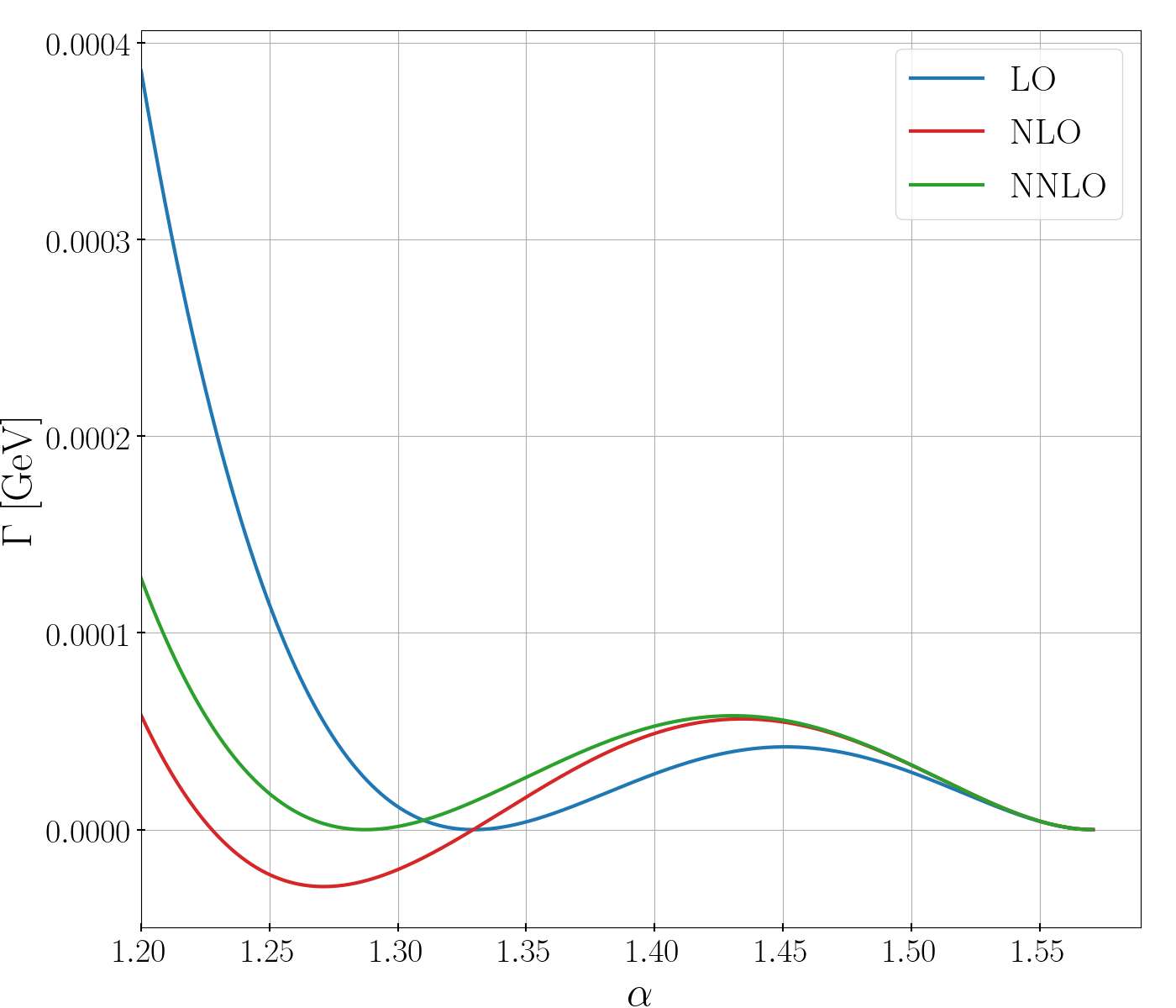}
    \caption{Partial decay width $\Gamma(h_2 \to h_1 h_1)$ for the
      benchmark point {\tt BPapprox} as a function of the mixing angle
      $\alpha$, at LO (blue), NLO (orange), and approximate NNLO (green). \label{fig:nnloapprox}} 
 \end{figure}

To illustrate this, we show in Fig.~\ref{fig:nnloapprox} for the benchmark point
\beq
\mbox{{\tt BPapprox: }} m_{h_1} = 60 \mbox{ GeV}, \; 
m_{h_2} = 125 \mbox{ GeV}, \; m_A = 40 \mbox{ GeV}, \; v_S = 1 \mbox{ TeV},
\eeq
and varying mixing angle $\alpha$ the decay width $\Gamma (h_2 \to h_1
h_1)$ at LO, NLO and approximate NNLO given by Eq.~(\ref{eq:partialnnlo}). For
$\alpha = \arctan(v_S/v) = 1.33$ the LO and NLO widths vanish, and the
result is given by the NNLO decay width, which is exact here and amounts to the 
value of $1.8 \times 10^{-5}$~GeV. For $\alpha > 1.33$ the LO
and NLO widths become non-zero with the NLO width gaining in
importance compared to the approximate NNLO result\footnote{For
  some region $\alpha < 1.33$ the NLO width becomes negative and hence
  unphysical as in this region the NLO amplitude is larger than the LO amplitude and with an opposite sign.}. Only at $\alpha =
1.388$, however, the NNLO width amounts to less than 10\% of the NLO
result. In the relatively large transition region $\alpha \in
[1.33,1.388]$ it is totally unclear, however, if the large difference
between NNLO and NLO is due to incomplete cancellations in the
approximate NNLO width and/or due to a small NLO width close to its
vanishing point at $\alpha=1.33$. Note, finally, that as expected all
amplitudes vanish in the SM-like limit which is obtained for $\alpha = \pi/2$.

\section{Implementation in sHDECAY: EWsHDECAY}
\label{sec:hdecay}
We implemented the EW one-loop corrections to the Higgs decays derived
in this work in the singlet extension {\tt sHDECAY} of the code {\tt
  HDECAY} \cite{Djouadi_1998,Djouadi_2019} where we updated the
underlying {\tt HDECAY} version to 
version 6.61. The resulting code is called {\tt EWsHDECAY}.
Note, that we impose an extra $\mathbb{Z}_2$ symmetry on
$S$ in contrast to the CxSM version implemented in {\tt sHDECAY}
\cite{shdecaywebpage}. In order to use our version of the CxSM
therefore the input parameter $a_1$ has to be set to zero in 
the input file, {\it cf.}~also footnote \ref{footnote1}. The Fortran code {\tt
  HDECAY} and thereby its extension {\tt sHDECAY} include the
state-of-the-art QCD corrections to the partial decay widths. For the
consistent combination of our EW corrections with {\tt HDECAY} which
uses the Fermi constant $G_F$ as input parameter, we use the $G_\mu$,
respectively the $G_F$ scheme, in the definition of our counterterm
for the electric charge, {\it cf.}~Subsection~\ref{sec:gaugesec}. \s

Note, that {\tt sHDECAY} also calculates the off-shell decays into
final states with an off-shell top-quark $t^*$, $\phi \to t^* \bar{t}$
($\phi=h_1,h_2$) and into final states with off-shell gauge bosons $V^*$
($V=W,Z$), $\phi \to V^* V^*$. The EW corrections are only computed
for on-shell decays, however. This means that if only off-shell decays
are kinematically allowed for the mentioned final states, then the EW
corrections are not included and only the LO decay width is calculated. Be
reminded also, that we did not include EW corrections to the loop-induced
decays into photonic and gluonic final states. We furthermore assume
that the QCD and EW corrections factorize. The relative QCD
corrections $\delta^{\text{QCD}}$ are defined relative to the LO width
$\Gamma^{\text{CxSM,LO}}$ calculated by {\tt sHDECAY}, which contains
(where applicable) running quark masses in order to improve the
perturbative behavior. The relative EW corrections
$\delta^{\text{EW}}$ on the other hand, are obtained by normalizing to
the LO width with OS particle masses. The QCD and EW corrected  decay
width into a specific OS and non-loop-induced final state,
$\Gamma^{\text{QCD\&EW}}$ is hence obtained as
\beq
\Gamma^{\text{QCD\&EW}} = \Gamma^{\text{CxSM,LO}}
[1+\delta^{\text{QCD}}] [1+\delta^{\text{EW}}] =
\Gamma^{\text{CxSM,QCD}} [1+\delta^{\text{EW}}] \;.
\label{eq:hodef}
\eeq
And the branching ratio of the higher-order (HO) corrected decay width
$h_i$ ($i=1,2$) into a specific final state is calculated by
\beq
\mbox{BR} (h_i \to XX) = \frac{\Gamma^{\text{HO}} (h_i \to
  XX)}{\Gamma_{\text{tot}}^{\text{QCD\&EW}}} \;,
\label{eq:bran}
\eeq
with the total width given by,
\beq
\Gamma_{\text{tot}}^{\text{QCD\&EW}} (h_i) &=& 
\Gamma^{\text{QCD}} (h_i \to gg) + \Gamma (h_i \to \gamma\gamma)
\nonumber \\
&& + \sum_{f=s,c,b,t} \Gamma^{\text{QCD\&EW}} (h_i \to ff) + 
\Gamma^{\text{EW}} (h_i \to \tau\tau) \label{eq:gamtot} \\
&& + 
\sum_{V=Z,W} \Gamma^{\text{EW}} (h_i \to VV) + 
\delta_{i2} \Gamma^{\text{EW}} (h_2 \to h_1 h_1) + 
\Gamma^{\text{EW}} (h_i \to AA) \;. \nonumber 
\eeq
As mentioned above, the QCD corrections to the decays into colored
final states are those 
as implemented in the original {\tt HDECAY} version, for details see
\cite{Djouadi_2019}. As mentioned above, the EW corrections are only included for
non-loop-induced final states and for on-shell decays. Otherwise the
off-shell tree-level decays 
are used if implemented in {\tt sHDECAY}, as is the case for the $tt$,
$ZZ$ and $WW$ final states, hence
\beq
\Gamma^{\text{EW}} (h_i \to XX) = \left\{ 
\begin{array}{lcl}
\Gamma^{\text{EW}} (h_i \to XX) & \mbox{for} & m_{h_i} \ge 2 m_X
  \\
\Gamma^{\text{LO}} (h_i \to X^* X^{*}) & \mbox{for} & m_{h_i} < 2
                                                       m_X 
\end{array}
\right. \;,
\quad \mbox{with } X=t,Z,W \;, \label{eq:partial1}
\eeq
and 
\beq
\Gamma^{\text{EW}} (h_i \to XX) = \left\{ 
\begin{array}{lcl}
\Gamma^{\text{EW}} (h_i \to XX) & \mbox{for} & m_{h_i} \ge 2 m_X
  \\
0 & \mbox{for} & m_{h_i} < 2 m_X
\end{array}
\right. \;,
\quad \mbox{with } X=b,\tau,h_1,A \;. \label{eq:partial2}
\eeq
Note, that for the fermionic final states we included the EW
corrections only of the third generation. As the decay widths in
fermion pairs of the 
first two generations are much smaller, the effect of not including
the EW corrections here is negligible.
The partial widths $\Gamma^{\text{HO}} (h_i \to XX)$ in
Eq.~(\ref{eq:bran}) are given as defined in Eqs.~(\ref{eq:hodef}) to
(\ref{eq:partial2}). \s

The parameters of the model are set in the corresponding block of the input file {\tt hdecay.in}
as in the original input file for {\tt sHDECAY}. 
\begin{verbatim}
********************** real or complex singlet Model *********************
Singlet Extension: 1 - yes, 0 - no (based on SM w/o EW corrections)
Model: 1 - real broken phase, 2 - real dark matter phase
       3 - complex broken phase, 4 - complex dark matter phase
isinglet = 1
icxSM    = 4
...
*** complex singlet dark matter phase ***
alph1    = -0.14331719167196D0
m1       = 125.09D0
m2       = 345.0021536863908D0
m3       = 61.02910887468706D0
vs       = 467.196443135993D0
a1       = 0.D0
\end{verbatim}
The model for which we compute the EW corrections is the CxSM in the
DM phase, so that the user has to choose 'icxSM=4' after
setting 'isinglet=1'. The input values are then given in the block
named 'complex singlet dark matter phase'. We consider a specific
version of the model where the parameter $a_1$ does not appear in the
model, so that it has to be set to 0 for the computation of the EW
corrections. Therefore, if the parameter is chosen to be non-zero, no EW
corrections will be calculated. Note, that $m_{1,2}$ denote the lighter and heavier of
the two visible Higgs boson masses, and $m_3$ is the mass value of the
DM particle. For the inclusion of the EW corrections the input file {\tt
  hdecay.in} of {\tt sHDECAY} has been extended by the following lines:
\begin{verbatim}
******************** EW Corrections ********************
** Attention: This can only be used for the complex dark matter phase
(icxSM=4) of the CxSM with a1 = 0 **
**** ielwcxsm = 0 LO, = 1 include NLO corrections
ielwcxsm = 1
**** ren. scheme: vsscheme = 1 pd, = 2 ZEM, pdprocess = 1 h1->AA, =2
h2->AA, alpha_mix =1 OS, =2 pstar
vsscheme = 2
pdprocess= 2
ralph_mix= 1
**** IR parameter - DeltaE is the detector resolution (in GeV)
DeltaE   = 10.0D0
**** NNLO approx: NNLOapprox=1, add NLO^2 term to h2h1h1 decay width 
if |tan(alpha)*(v/vs)-1|<deltaNNLO
NNLOapp  = 1
deltaNNLO= 0.05D0
**** Parameter conversion, change input parameters vS and alpha
accordingly, if given scheme above is not the specified input scheme
**** Paramcon = 0 no parameter conversion, =1 do parameter conversion
**** Standard scheme: stdvs = 1 pd, = 2 ZEM, stdproc = 1 h1->AA, 
=2 h2->AA, stdalpha =1 OS, =2 pstar
Paramcon = 0
stdvs    = 1
stdproc  = 2
stdalpha = 1
\end{verbatim}
The warning reminds the user that our computation of the EW
corrections only applies to the CxSM in the DM phase with the
$a_1=0$ in the potential. By setting the input parameter
'ielwcxsm' equal to 1 (0), the EW corrections will be computed
(omitted). The next three parameters choose the applied
renormalization scheme. Setting 'vsscheme' to 1, the singlet VEV
counterterm will be computed in the process-dependent scheme using the
on-shell decay $h_1 \to AA$ ($h_2 \to AA$) by choosing 'pdprocess =1
(2)' (called 'OSproc1' and 'OSproc2', respectively, in
Tab.~\ref{tab:renscheme}). When 'vsscheme' is set to 2, then the process-dependent
renormalization scheme for $v_s$ is evaluated at zero external
momenta for the chosen decay process (called 'ZEMproc1' and 'ZEMproc2',
respectively in Tab.~\ref{tab:renscheme}). With 'ralph\_mix=1 (2)' the
OS-pinched ($p_*$-pinched) scheme is chosen for the renormalization of the mixing
angle $\alpha$. \s 

The next parameter setting chooses with 'DeltaE' 
the detector resolution needed in the computation of the IR
corrections, which we set by default equal to 
10~GeV, {\it cf.}~\cite{Krause:2016oke}. The following two parameters
concern the computation of the NNLO corrections to the decay width $h_2 \to
h_1 h_1$. If 'NNLOapp' is set equal to 1 then the NNLO width is
computed for parameter configurations with $\tan\alpha$ values in the
vicinity of $v_S/v$, namely $\tan 
\alpha \in \{v_s/v - \mbox{deltaNNLO}, v_s/v + \mbox{deltaNNLO}\}$. Be
aware, however, that for $\tan\alpha \ne v_s/v$, the NNLO computation is
incomplete, as discussed above. \s

The last four parameters refer to the
  conversion of the input parameters. The change of the
  renormalization scheme allows for an estimate of the uncertainty in
  the NLO EW corrections due to the missing higher-order
  corrections. For such an estimate to be meaningful also the input parameters have to
  be changed consistently. For the conversion of the input
parameters, when going from one to the other renormalization scheme, our code uses an approximate formula based on the bare parameter
$p_0$ which is independent of the scheme so that,
\beq
p_{\text{spec}} = p_{\text{std}} + \delta p_{\text{std}} - \delta
p_{\text{spec}} \;, \label{eq:schemechange}
\eeq
where $p_x$ and $\delta p_x$ denote the renormalized parameter $p$ and
its counterterm $\delta p$ in the standard scheme and in the user specified scheme,
respectively. Setting the flag ' Paramcon = 1', the parameter conversion is 
applied. In this case, the user can then also choose which
renormalization scheme is the 'standard' scheme, by setting the 
three parameters '(stdvs,stdproc,stdalpha)'. If the user specified
renormalization scheme '(vsscheme,pdprocess,ralph)' and 
the standard scheme '(stdvs,stdproc,stdalpha)' are identical, the
input parameters $v_S$ and $\alpha$ remain
unchanged. Otherwise they are converted from their values understood
to be given in the standard scheme '(stdvs,stdproc,stdalpha)' to the
values they take in the user specified scheme
'(vsscheme,pdprocess,ralph)'. The converted parameter values are given
out in the file 'Paramconversion.txt'. If, however, the user chooses 'Paramcon = 0' the
parameters are not converted when the specified renormalization scheme
differs from the standard scheme. In this case a comparison of the results for different
renormalization schemes to estimate the theoretical uncertainty would
not, however, be meaningful. \s

Note that we do not give out the NLO value 
of a specific decay width if it becomes negative\footnote{This is the
  case {\it e.g.}~close to a vanishing LO decay width that is
  proportional to a small coupling
  squared, whereas the NLO width only linearly depends on this
  coupling.}. In this case, and also if a chosen renormalization scheme
cannot be applied because it is kinematically not allowed, the decay
width is computed and given out at LO in the EW corrections together
with a warning. Remark, finally, that when we take the SM limit of our
model and compare the results for the EW corrections to those obtained
from {\tt HDECAY} for the SM, we differ in the decay into gauge boson
final states as {\tt HDECAY} includes the EW corrections also to the
off-shell SM Higgs decays into $WW/ZZ$ and to the loop-induced decays
into gluon and photon pairs. \s

The code can be
downloaded at the url: \\
\centerline{\url{https://github.com/fegle/ewshdecay}}
\\
Apart from short explanations and user instructions, we also provide
sample input and output files on this webpage. 

\section{Parameter Scan}
\label{sec:scan}
In order to obtain viable parameter points for our numerical analysis
we performed a scan in the parameter space of the model and kept only
those points which fulfill the relevant theoretical and experimental
constraints described below. They are checked for by {\tt ScannerS}
\cite{Coimbra:2013qq,Muhlleitner:2020wwk} which we used for the
parameter scan, with the scan ranges summarized in Tab.~\ref{tab:scanranges}.
\begin{table}[h!]
\centering
\begin{tabular}{ccc}
\toprule
Parameter & \multicolumn{2}{c}{Range} \\
\cmidrule{2-3}
 & {Lower} & {Upper} \\
\midrule
$m_s$ & 30 GeV & 1000 GeV\\
$m_A$ & 10 GeV & 1000 GeV \\
$v_S$ & 1 GeV & 1000 GeV\\
$\alpha$ & $-$1.57 &  1.57 \\
\bottomrule
\end{tabular}
\caption{The scan ranges used for the generation of parameter points
  with \texttt{ScannerS}. The mass $m_s$ denotes the scalar mass of
  the non-125~GeV Higgs boson. 
\label{tab:scanranges}}
\end{table}
The values of the SM input parameters needed for our calculation are given in Tab. \ref{tab:inputparameters}. These are the
values suggested by the LHC Higgs Working Group \cite{LHCHXSWG}. \s
\begin{table}[h!]
\centering
\begin{tabular}{cc}
\toprule
SM parameter & Value \\
\midrule
$m_Z$ & 91.15348 GeV \\
$m_W$ & 80.3579 GeV\\
$m_{h_{125}}$ & 125.09 GeV\\
$m_\tau$ & 1.77682 GeV\\
$\overline{m}_b (m_b)$ & 4.18 GeV  \\
$m_t$ & 172.5 GeV\\
\bottomrule
\end{tabular}
\caption{The SM parameter values used in the numerical evaluation
  taken from \cite{Zyla:2020zbs} and following the recommendations of
  \cite{LHCHXSWG}. Here, $\overline{m}_b$ denotes the 
  $\overline{\mbox{MS}}$ bottom quark mass which is taken at the scale
  of the bottom quark mass, $m_b$, and $h_{125}$ denotes the SM-like
  Higgs boson with a mass of 125.09~GeV.
  }
\label{tab:inputparameters}
\end{table}

The theoretical constraints that are taken into account are
boundedness from below, perturbative unitarity\footnote{We require the
eigenvalues of the scattering matrix of all possible two-to-two scalar
scatterings to be below $8\pi$ \cite{Lee:1977eg}.} and stability of the
vacuum. For detailed information, we refer to \cite{Egle:2022wmq}. \s

As for the experimental constraints, we first note that in our model
the $\rho$ parameter is equal to 1 at tree level and there are no
tree-level flavour-changing neutral currents, as the gauge singlet
does not couple to fermions and gauge bosons in the gauge basis. We
check for compatibility with EW precision data by requiring the
$S,T,U$ parameters \cite{PhysRevD.46.381} to be consistent with the
measured quantities at 95\% confidence level. Compatibility with the
LHC Higgs data and exclusion bounds is checked through the {\tt
  ScannerS} link with {\tt HiggsSignals}
\cite{Bechtle:2013xfa,Bechtle:2020uwn} and {\tt HiggsBounds}
\cite{Bechtle:2008jh,Bechtle:2020pkv}. We  
require the signal rates of our SM-like Higgs boson to be in agreement
with the experimental data at the $2\sigma$ level. Through the link with
{\tt MicrOmegas} \cite{Belanger:2006is} we ensure that the DM relic
density of our model does not exceed the measured value.  Concerning DM
direct detection, the tree-level cross section is negligible in this
model \cite{Gross:2017dan,Azevedo:2018oxv}, but the loop-corrected
DM-nucleon spin-independent cross section
\cite{Azevedo:2018exj,Glaus:2020ihj} has to be checked to be below
the experimental bounds \cite{XENON:2018voc,PandaX-4T:2021bab,LUXExperiment} with the most stringent bounds given by the LUX-ZEPLIN experiment
\cite{LUXExperiment}. Further detailed information on the experimental
constraints that we applied is given in \cite{Egle:2022wmq}. \s 

In our sample we take off parameter scenarios where the deviation of
any other neutral scalar mass from the $h_{125}$ mass is below
2.5~GeV. This suppresses 
interfering di-Higgs signals which would require a further special
treatment to correctly consider theory and experiment beyond the scope
and focus of the present work. \s
 
We also used the program {\tt BSMPT}
\cite{Basler:2018cwe,Basler:2020nrq} to check if the thus obtained
parameter sample leads to vacuum states compatible with an EW
VEV $v=246$~GeV also at NLO, {\it i.e.}~after including higher-order
corrections to the effective potential. Provided this to be the case,
we checked if we have points providing a strong first-order EW phase
transition (SFOEWPT), one of the Sakharov conditions
\cite{Sakharov:1967dj} to be fulfilled for EW 
baryogenesis. We found that none of the parameter points with a viable
NLO EW vacuum provides an SFOEWPT. \s

Finally, we applied constraints from the resonant di-Higgs searches
performed by the LHC experiments. We proceed here as described in
\cite{Abouabid:2021yvw} to which we also refer for a detailed
description, giving here only the most important points. In order to
apply the constraints, we calculated for parameter points of scenario I, where $m_{h_2} >
2 m_{h_1} \equiv 2 m_{h_{125}}$ the production cross section $\sigma
(h_2)$ at next-to-next-to-leading-order QCD using the code {\tt SusHi v1.6.1}
\cite{Harlander:2012pb,Liebler:2015bka,Harlander:2016hcx} and
multiplied it with the branching ration $\mbox{BR} (h_2 \to h_1
h_1)$. We then checked if the thus obtained rate is below the
experimental values given in Refs.~\cite{ATLAS:2018rnh, ATLAS-CONF-2021-035, CMS-PAS-B2G-20-004} for the $4b$,
\cite{ATLAS:2018uni,ATLAS:2020azv, ATLAS-CONF-2021-030} for the $(2b)(2\tau)$,
\cite{ATLAS:2018dpp, ATLAS-CONF-2021-016} for the $(2b)(2\gamma)$,
\cite{ATLAS:2018fpd} for 
the $(2b)(2W)$, \cite{CMS:2020jeo} for the $(2b)(ZZ)$, \cite{ATLAS:2018hqk} for the
$(2W)(2\gamma)$ and \cite{ATLAS:2018ili} for the $4W$ final
states. There are also experimental limits from non-resonant
searches. They do not constrain our model so far, however, as will be
discussed below. \s

The parameter points which are found to be compatible with all applied
constraints can be divided up into two samples, depending on which of
the visible Higgs particles is the SM-like one, called $h_{125}$ in
the following. For the two samples, we will adopt the following
notation,
\beq
\begin{array}{ll}
m_{h_1} = m_{h_{125}}: & \mbox{ scenario I} \\
m_{h_2} = m_{h_{125}}: & \mbox{ scenario II} \;.
\end{array}
\eeq

\section{Numerical Analysis \label{sec:numerical}}
 We start our numerical analysis by investigating what is the impact of
our extended scan compared to \cite{Egle:2022wmq} on the allowed
parameter regions of the model. We move on to the discussion of
the overall size of the computed EW corrections before investigating
the remaining theoretical uncertainty due to missing higher-order
corrections. We then discuss the 
phenomenological impact of these corrections w.r.t~the Higgs decays
into DM particles and the impact on the allowed parameter
regions. Finally, we investigate Higgs pair production in the context
of our model and how it is affected by our corrections.

\subsection{Allowed Parameter Regions and DM Observables}
\begin{figure}[t!]
    \centering
    \includegraphics[width=0.8\textwidth]{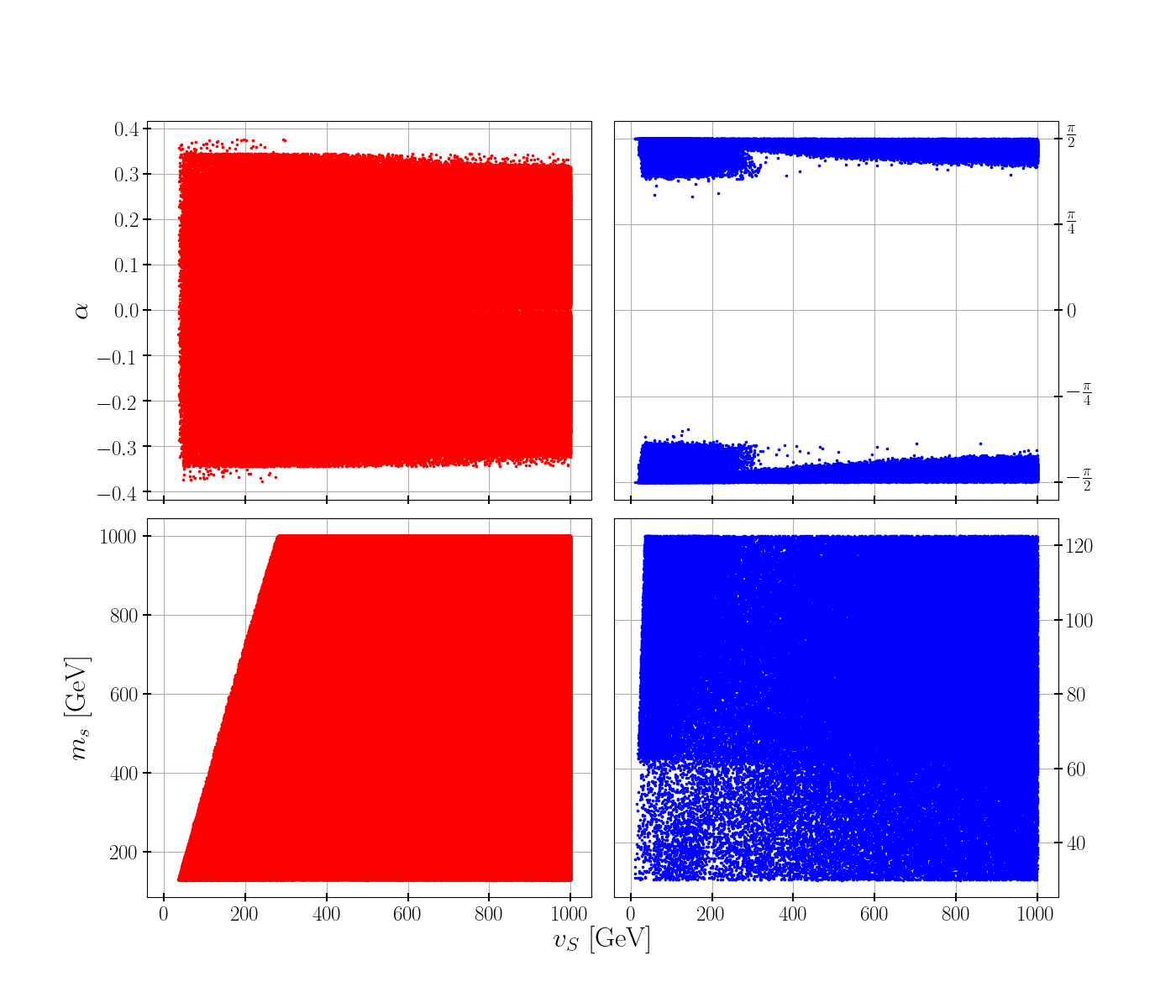}
\vspace*{-0.8cm}
   \caption{Allowed input parameters $\alpha$ vs.~$v_S$ (upper row) and $m_S$
      vs.~$v_S$ (lower row). Left: for scenario I ($h_1 = h_{125}$,
      red points); right: for scenario II ($h_2 = h_{125}$, blue points).\label{fig:parvalues}}
\end{figure}

In our new scan we go up to $m_A$ values of 1~TeV, as compared to \cite{Egle:2022wmq}, we now also allow for
scenarios where the decay of the SM-like Higgs boson into DM particles
is kinematically closed. This leads to an enlarged parameter space, a
larger variation in the particle couplings and masses and hence
effects in the NLO EW corrections that would not appear in the more
restricted sample. We therefore show the corresponding plot to Fig.~1
in \cite{Egle:2022wmq}. Figure~\ref{fig:parvalues} (upper) displays
the allowed combinations of $\alpha$ and $v_S$ for scenario I,
{\it i.e.}~$h_1 = h_{125}$, (left) and for scenario II, {\it i.e.}~$h_2 =
h_{125}$, (right).\footnote{Unless stated otherwise, all plots shown
  here and in the following are obtained for values that are checked
  for Higgs constraints by applying the widths at LO in the EW corrections.} The lower plot shows the corresponding values of 
non-125-GeV Higgs boson mass $m_S$ vs.~$v_S$. Compared to
\cite{Egle:2022wmq}, we see that $\alpha$ and $v_S$ 
are not linked any more as for the large DM matter masses, that are now
included in our scan, the kinematic constraints inducing this
relation are not applicable any more. This leads also to 
a larger allowed maximum range for $|\alpha|$ which has increased in
scenario I from about 0.27 to about 0.34 for almost all parameter
points of scenario I. This is the maximal allowed value compatible
with the Higgs data except for parameter points where the visible
Higgs bosons are close in mass and the Higgs signal is built up by the
two Higgs bosons. These are the outliers with larger mixing angles up
to about 0.37. If we dropped the exclusion of parameter points with
non-SM-Higgs masses closer than 2.5~GeV to the SM-like Higgs mass then
the mixing could be even larger.\footnote{The fact that these points appear for
singlet VEVs $v_S \lsim 300$~GeV, is related to the relic density
constraints. Larger $v_S$ values combined with nearly degenerate Higgs
boson masses induce small Higgs portal couplings and hence
increased relic densities.}
Also in scenario II larger deviations from $\alpha= \pm  
\pi/2$ are possible for the increased parameter scan.\footnote{Note that the kink in the distribution
  of points of scenario II at $v_S \approx 280$~GeV is simply a scan
  artifact stemming from the fact that we performed an additional
  dedicated scan for larger $m_A$ values.} The
lower plots, apart from becoming more dense, did not
change the shape. The perturbative unitarity constraints are reflected
in the linear relation between $m_S$ and $v_S$, {\it cf.}~the
discussion in \cite{Egle:2022wmq}. \s

\begin{figure}[t!]
    \centering
    \includegraphics[width=0.6\textwidth]{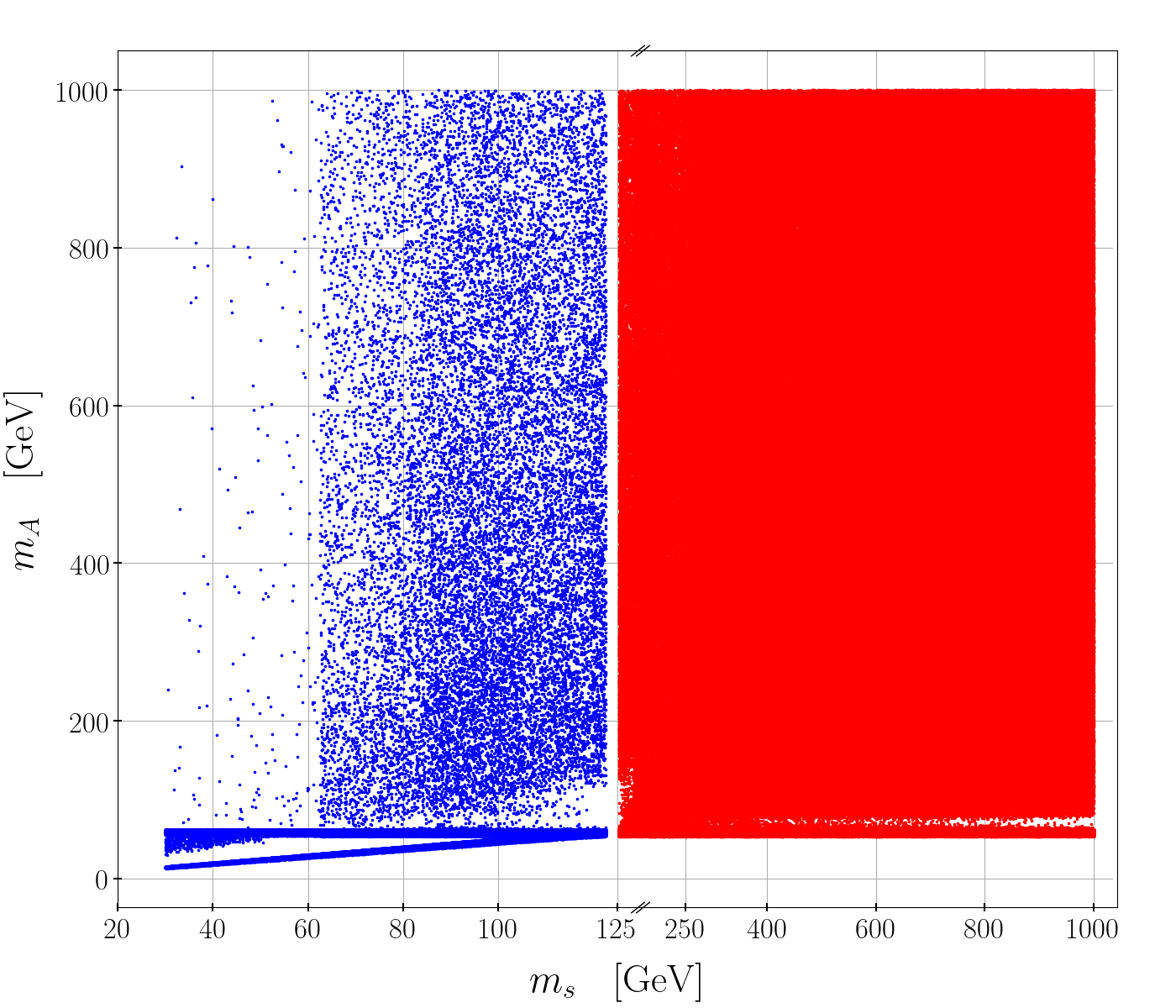}
\vspace*{-0.4cm}
    \caption{Allowed DM mass values $m_A$ versus non-SM-like Higgs mass
      values $m_S$ in GeV for scenario II ($h_2 = h_{125}$, blue
      points)  and scenario I ($h_1 = h_{125}$, red points).\label{fig:massvalues}}
 \end{figure}

Figure~\ref{fig:massvalues} displays the allowed values of the DM mass
$m_A$ versus the non-SM-like Higgs mass mass $m_S$ for scenario I (red
points) and scenario II (blue points). For $m_A$ values below
62.5~GeV, only mass values in the vicinity of $m_{h_{125}}/2$ or
$m_S/2$ are allowed, as then efficient DM annihilation via $h_{125}$
or the non-SM-like Higgs particle can take place such that the DM
relic density constraints can be met ({\it cf.}~our discussion in
\cite{Egle:2022wmq}). For large $m_A$ values where annihilation via
other SM particles can take place all mass values up to
the upper limit of our scan are allowed. \s

\begin{figure}[t!]
    \centering
    \includegraphics[width=1.0\textwidth]{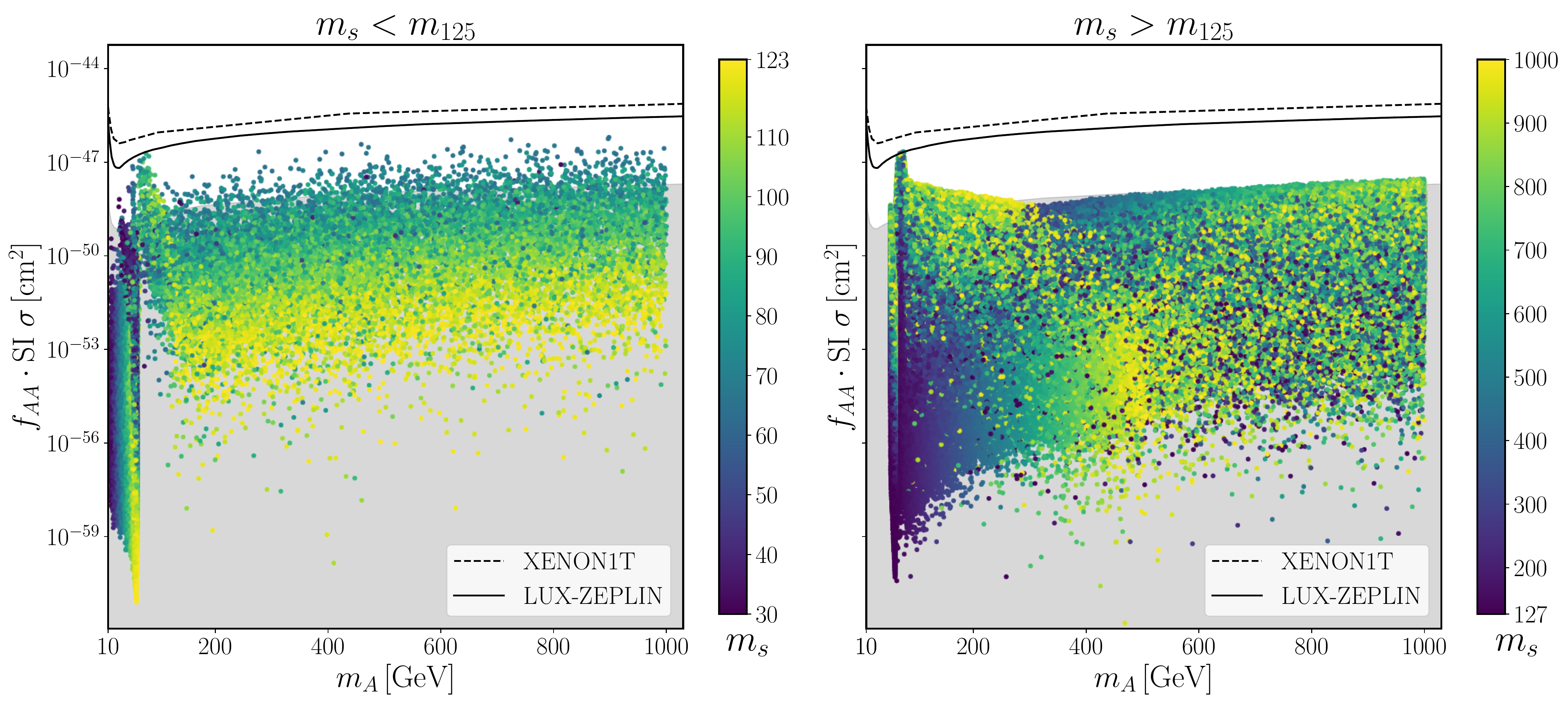}
\vspace*{-0.6cm}  
  \caption{Effective spin-independent nucleon DM cross section as a
      function of the DM mass for scenario II (left) and scenario I
      (right). The limits from XENON1T \cite{XENON:2018voc} and
      LUX-ZEPLIN \cite{LUXExperiment} 
      are given by the dashed and full line, respectively. The grey
      shaded region corresponds to the neutrino floor. The color code
      represents the non-SM-like 
      Higgs mass value in GeV.\label{fig:dirdetec}}
 \end{figure} 

In Fig.~\ref{fig:dirdetec} we show the effective spin-independent
nucleon DM detection cross section as a function of the DM mass $m_A$
for scenario II (left) and scenario I (right). The multiplication of
the spin-independent cross section SI $\sigma$ with the factor
\beq
f_{AA} = \frac{(\Omega h^2)_A}{(\Omega h^2)^{\text{obs}}_{\text{DM}}}
\eeq 
accounts for the fact that the relic density $(\Omega h^2)_A$ of our DM
candidate $A$ might not account for the entire observed DM relic
density $(\Omega h^2)^{\text{obs}}_{\text{DM}} = 0.120 \pm 0.001$
\cite{Aghanim:2018eyx}.\footnote{We find, 
  however, in our scan that for any DM mass that is still allowed
  there are parameter points that saturate the relic density.} In our model the
direct detection cross section at LO is negligible due to a
cancellation \cite{Gross:2017dan,Azevedo:2018oxv} so that we present
the one-loop result calculated in 
\cite{Azevedo:2018exj, Glaus:2020ihj}. In \cite{Egle:2022wmq}, we made
a mistake that we correct 
here so that now both the exclusion limits from Xenon1T \cite{XENON:2018voc}
(dashed) and LUX-ZEPLIN \cite{LUXExperiment} (full) as well as the neutrino floor (grey
region) move down by one order of magnitude. The effect is that for
both scenarios I and II we have now parameter points above the neutrino floor
that can hence be tested by direct detection experiments. Moreover, we
find that in the parameter region $m_A \ge m_{h_{125}}/2$, where we
cannot probe DM at the LHC,  the LUX-ZEPLIN experiment
\cite{LUXExperiment} is sensitive to  
the model in the region $66 \mbox{ GeV } \lsim m_A \lsim 78$~GeV. Future increased
precision in the direct detection experiments will allow to test the
model in (large) parts of the still allowed $m_A$ range in scenario I
(scenario II). \s

\begin{figure}[h!]
    \centering
    \includegraphics[width=1\textwidth]{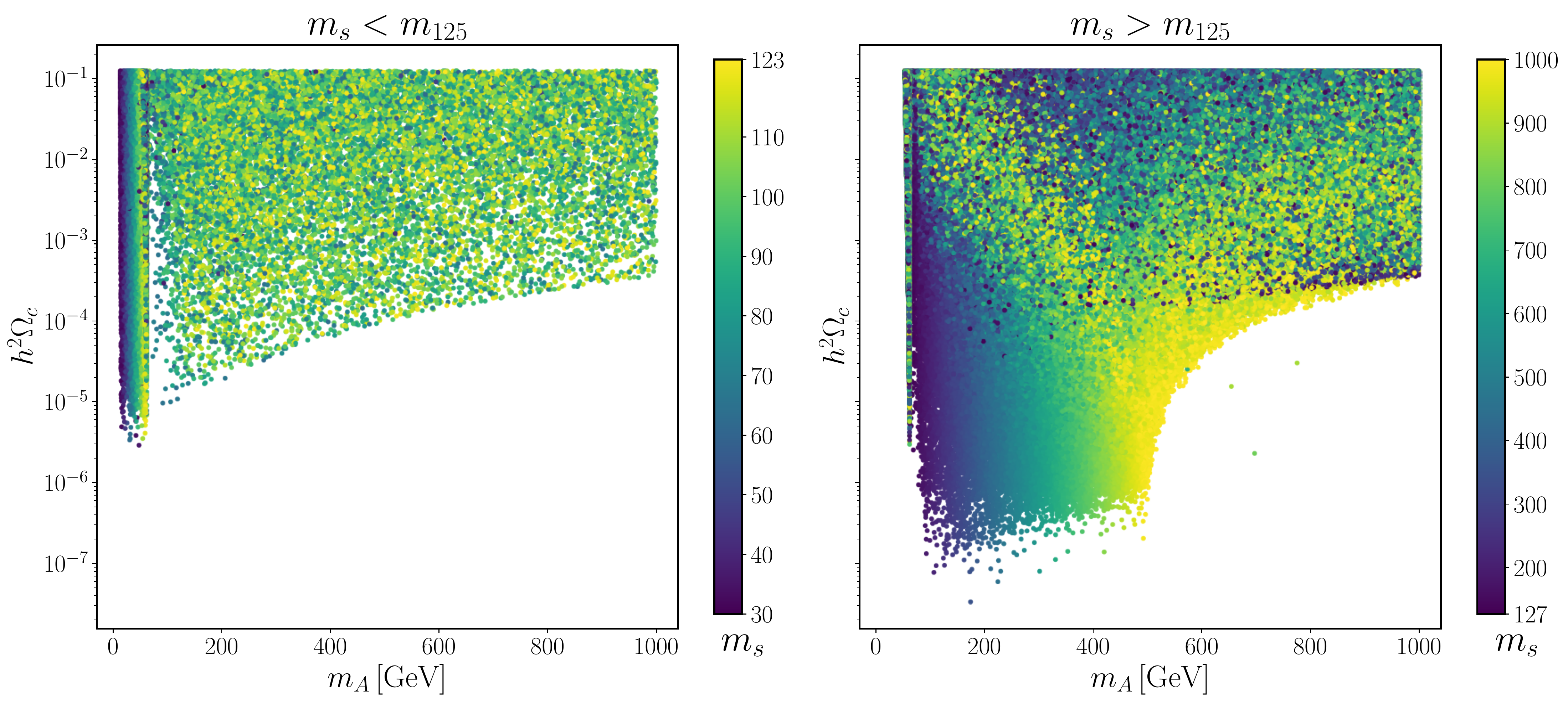}
\vspace*{-0.6cm}
    \caption{Relic density as a function of the DM mass $m_A$ for both
      II (left) and scenario I (right). The color code represents the non-SM-like
      Higgs mass value in GeV. \label{fig:relicdensity}} 
 \end{figure} 

Figure \ref{fig:relicdensity} displays the relic density for all
points passing our constraints for scenario II (left) and scenario I
(right). As can be inferred from the plot for 
the whole allowed region up to $m_A = 1$~TeV, where our scan ends, there are
parameter scenarios that saturate the relic density. 

\subsection{Size of NLO EW Corrections \label{subsec:size}}
\begin{table}[h!]
\centering
\begin{tabular}{|c||c|c||c|c||c|c||c|c|}
\hline
& \multicolumn{2}{c||}{OSproc1-OS}& \multicolumn{2}{c||}{OSproc1-$p_*$}& \multicolumn{2}{c||}{OSproc2-OS}& \multicolumn{2}{c|}{OSproc2-$p_*$}\\
 & {Min} & {Max} & {Min} & {Max} & {Min} & {Max} & {Min} & {Max}\\
\hline
$h_1 \to bb$ &\SI{-1.45}{\percent}&\SI{0.0}{\percent}&\SI{-1.41}{\percent}&\SI{0.0}{\percent}&\SI{-4.66}{\percent}&\SI{0.0}{\percent}&\SI{-4.79}{\percent}&\SI{0.0}{\percent}\\
$h_1\to \tau \tau $ &\SI{-4.65}{\percent}&\SI{0.0}{\percent}&\SI{-4.61}{\percent}&\SI{0.0}{\percent}&\SI{-7.85}{\percent}&\SI{0.0}{\percent}&\SI{-7.99}{\percent}&\SI{0.0}{\percent}\\
$h_1 \to AA$ &\SI{0}{\percent}&\SI{0}{\percent}&\SI{0}{\percent}&\SI{0}{\percent}&\SI{-2.12}{\percent}&\SI{6.78}{\percent}&\SI{-2.24}{\percent}&\SI{7.18}{\percent}\\
$h_2 \to bb$ &\SI{-9.19}{\percent}&\SI{4.85}{\percent}&\SI{-9.57}{\percent}&\SI{5.26}{\percent}&\SI{-10.18}{\percent}&\SI{19.72}{\percent}&\SI{-10.52}{\percent}&\SI{25.06}{\percent}\\
$h_2 \to \tau \tau$ &\SI{-16.3}{\percent}&\SI{0.0}{\percent}&\SI{-16.79}{\percent}&\SI{0.0}{\percent}&\SI{-19.03}{\percent}&\SI{16.39}{\percent}&\SI{-19.5}{\percent}&\SI{21.74}{\percent}\\
$h_2 \to tt$ &\SI{-6.17}{\percent}&\SI{5.47}{\percent}&\SI{-6.17}{\percent}&\SI{6.13}{\percent}&\SI{-6.16}{\percent}&\SI{11.05}{\percent}&\SI{-6.16}{\percent}&\SI{10.93}{\percent}\\
$h_2 \to WW$ &\SI{-9.82}{\percent}&\SI{9.31}{\percent}&\SI{-9.85}{\percent}&\SI{9.5}{\percent}&\SI{-9.8}{\percent}&\SI{19.49}{\percent}&\SI{-9.89}{\percent}&\SI{21.53}{\percent}\\
$h_2 \to ZZ$ &\SI{-3.53}{\percent}&\SI{4.44}{\percent}&\SI{-3.51}{\percent}&\SI{5.27}{\percent}&\SI{-3.54}{\percent}&\SI{12.1}{\percent}&\SI{-3.59}{\percent}&\SI{13.33}{\percent}\\
$h_2 \to h_1 h_1$ &\SI{-14.14}{\percent}&\SI{10.64}{\percent}&\SI{-14.04}{\percent}&\SI{10.54}{\percent}& \SI{-99.4}{\percent}& $>$\SI{100}{\percent}& \SI{-98.24}{\percent}& $>$\SI{100}{\percent}\\
$h_2 \to AA$ &\SI{-6.75}{\percent}&\SI{2.14}{\percent}&\SI{-7.2}{\percent}&\SI{2.25}{\percent}&\SI{0}{\percent}&\SI{0}{\percent}&\SI{0}{\percent}&\SI{0}{\percent}\\
\hline
\hline
& \multicolumn{2}{c||}{ZEMproc1-OS}& \multicolumn{2}{c||}{ZEMproc1-$p_*$}& \multicolumn{2}{c||}{ZEMproc2-OS}& \multicolumn{2}{c|}{ZEMproc2-$p_*$}\\
 & {Min} & {Max} & {Min} & {Max} & {Min} & {Max} & {Min} & {Max}\\
\hline
$h_1 \to bb$ &\SI{-4.66}{\percent}&\SI{-0.88}{\percent}&\SI{-4.79}{\percent}&\SI{-0.62}{\percent}&\SI{-4.66}{\percent}&\SI{-0.88}{\percent}&\SI{-4.79}{\percent}&\SI{-0.62}{\percent}\\
$h_1\to \tau \tau $ &\SI{-7.85}{\percent}&\SI{-4.08}{\percent}&\SI{-7.99}{\percent}&\SI{-3.8}{\percent}&\SI{-7.85}{\percent}&\SI{-4.08}{\percent}&\SI{-7.99}{\percent}&\SI{-3.8}{\percent}\\
$h_1 \to AA$ &\SI{-3.5}{\percent}&\SI{5.76}{\percent}&\SI{-3.5}{\percent}&\SI{5.76}{\percent}&\SI{-4.63}{\percent}&\SI{9.51}{\percent}&\SI{-5.0}{\percent}&\SI{9.93}{\percent}\\
$h_2 \to bb$ & \SI{-99.94}{\percent}&\SI{19.77}{\percent}& \SI{-99.57}{\percent}&\SI{25.04}{\percent}& \SI{-99.94}{\percent}&\SI{19.71}{\percent}& \SI{-99.57}{\percent}&\SI{25.06}{\percent}\\
$h_2 \to \tau \tau$ & \SI{-99.32}{\percent}&\SI{16.37}{\percent}& \SI{-98.96}{\percent}&\SI{21.7}{\percent}& \SI{-99.32}{\percent}&\SI{16.38}{\percent}& \SI{-98.96}{\percent}&\SI{21.73}{\percent}\\
$h_2 \to tt$ & \SI{-99.28}{\percent}&\SI{11.06}{\percent}& \SI{-99.11}{\percent}&\SI{10.93}{\percent}& \SI{-99.28}{\percent}&\SI{11.02}{\percent}& \SI{-99.11}{\percent}&\SI{10.91}{\percent}\\
$h_2 \to WW$ & \SI{-98.69}{\percent}&\SI{19.46}{\percent}& \SI{-98.81}{\percent}&\SI{21.55}{\percent}& \SI{-98.69}{\percent}&\SI{19.45}{\percent}& \SI{-98.81}{\percent}&\SI{21.52}{\percent}\\
$h_2 \to ZZ$ & \SI{-98.18}{\percent}&\SI{12.08}{\percent}& \SI{-97.8}{\percent}&\SI{13.29}{\percent}& \SI{-98.18}{\percent}&\SI{12.06}{\percent}& \SI{-97.8}{\percent}&\SI{13.32}{\percent}\\
$h_2 \to h_1 h_1$ & \SI{-99.95}{\percent}& $>$\SI{100}{\percent}& \SI{-99.96}{\percent}& $>$\SI{100}{\percent}& \SI{-99.92}{\percent}& $>$\SI{100}{\percent}& \SI{-99.88}{\percent}& $>$\SI{100}{\percent}\\
$h_2 \to AA$ & \SI{-100.0}{\percent}&\SI{5.65}{\percent}& \SI{-100.0}{\percent}&\SI{5.82}{\percent}&\SI{-15.22}{\percent}&\SI{9.03}{\percent}&\SI{-15.23}{\percent}&\SI{9.09}{\percent}\\
\hline
\end{tabular}
\caption{Scenario I: Relative NLO corrections $\delta_{\text{EW}} (h_i \to XX)$
  for all possible decay processes in 
  scenario I, {\it i.e.}~$m_{h_1}= 125.09$~GeV, and different renormalization
  schemes $\mbox{ren}_{v_S}$-$\mbox{ren}_\alpha$. For $\delta v_S$ we
  have OS renormalization and ZEM renormalization in the two possible processes,
  $\mbox{ren}_{v_S}=$ OSproc1, OSproc2, ZEMproc1, ZEMproc2. For $\delta \alpha$ we have
  the two possible schemes $\mbox{ren}_{\alpha}=$ OS, $p_*$. The 
  renormalization schemes and their notation are summarized in
  Tab.~\ref{tab:renscheme}. \label{tab:scenIall}}
\end{table}

In Tab.~\ref{tab:scenIall} we show the
relative NLO corrections to the partial widths $\Gamma$ of the decay of
$h_{i}$ ($i=1,2$) into the various possible final states $XX$
($X=\tau,b,t,W,Z,A,h_1$), defined as 
\beq
\delta_{\text{EW}} (h_i \to XX) = \frac{\Gamma^{\text{NLO}}_{h_i \to
    XX} - \Gamma^{\text{LO}}_{h_i \to XX}}{\Gamma^{\text{LO}}_{h_i \to
    XX}} \,, \label{eq:relewcorr}
\eeq
for scenario I (Tab.~\ref{tab:scenIall}) and scenario II
(Tab.~\ref{tab:scenIIall}) and different renormalization
schemes. These are for the renormalization of the singlet VEV $v_S$
the process-dependent and the ZEM scheme, using either the $h_1 \to
AA$ or $h_2 \to AA$ decay and for the renormalization of $\alpha$ the
OS- or the $p_\star$-pinched scheme. The notation for the schemes is summarized in
Tab.~\ref{tab:renscheme}. We also note that we discard all parameter points
where the large relative corrections are beyond -100\%, in order to not 
encounter unphysical results with negative decay widths at NLO. 
We furthermore emphasize that the results given in a specific
renormalization scheme assume the input parameters to be given in this
specific 
renormalization scheme. We hence do not start from a certain scheme $a$
and then move on to a different scheme $b$ by consistently converting the
input parameters of scheme $a$ to scheme $b$. The results given in the
various schemes hence have to be looked at and discussed each one by
one. They are meant to give an overview of what sizes of corrections
can be expected in the various schemes. 
\s

\begin{table}
\centering
\begin{tabular}{|c||c|c||c|c||c|c||c|c|}
\hline
& \multicolumn{2}{c||}{OSproc1-OS}& \multicolumn{2}{c||}{OSproc1-$p_*$}& \multicolumn{2}{c||}{OSproc2-OS}& \multicolumn{2}{c|}{OSproc2-$p_*$}\\
 & {Min} & {Max} & {Min} & {Max} & {Min} & {Max} & {Min} & {Max}\\
\hline
$h_1 \to bb$ &\SI{-1.37}{\percent}&\SI{6.02}{\percent}&\SI{-1.68}{\percent}&\SI{0.81}{\percent}& \SI{-49.18}{\percent}&\SI{27.16}{\percent}& \SI{-48.31}{\percent}& \SI{37.35}{\percent}\\
$h_1\to \tau \tau $ &\SI{-4.47}{\percent}&\SI{5.32}{\percent}&\SI{-4.48}{\percent}&\SI{1.68}{\percent}& \SI{-52.03}{\percent}&\SI{26.52}{\percent}& \SI{-51.16}{\percent}& \SI{37.53}{\percent}\\
$h_1 \to AA$ &\SI{0}{\percent}&\SI{0}{\percent}&\SI{0}{\percent}&\SI{0}{\percent}&\SI{-5.57}{\percent}&\SI{3.63}{\percent}&\SI{-1.24}{\percent}&\SI{5.16}{\percent}\\
$h_2 \to bb$ &\SI{-1.39}{\percent}&\SI{0.0}{\percent}&\SI{-1.39}{\percent}&\SI{0.0}{\percent}&\SI{-1.4}{\percent}&\SI{0.0}{\percent}&\SI{-1.4}{\percent}&\SI{0.0}{\percent}\\
$h_2 \to \tau \tau$ &\SI{-4.57}{\percent}&\SI{0.0}{\percent}&\SI{-4.59}{\percent}&\SI{0.0}{\percent}&\SI{-4.6}{\percent}&\SI{0.0}{\percent}&\SI{-4.59}{\percent}&\SI{0.0}{\percent}\\
$h_2 \to h_1 h_1$ &\SI{-10.6}{\percent}& \SI{46.49}{\percent}&\SI{-10.76}{\percent}& \SI{46.55}{\percent}&\SI{-11.64}{\percent}& $>$\SI{100}{\percent}&\SI{-11.7}{\percent}& $>$\SI{100}{\percent}\\
$h_2 \to AA$ &\SI{-3.64}{\percent}&\SI{5.6}{\percent}&\SI{-5.18}{\percent}&\SI{1.22}{\percent}&\SI{0}{\percent}&\SI{0}{\percent}&\SI{0}{\percent}&\SI{0}{\percent}\\
\hline
\hline
& \multicolumn{2}{c||}{ZEMproc1-OS}& \multicolumn{2}{c||}{ZEMproc1-$p_*$}& \multicolumn{2}{c||}{ZEMproc2-OS}& \multicolumn{2}{c|}{ZEMproc2-$p_*$}\\
 & {Min} & {Max} & {Min} & {Max} & {Min} & {Max} & {Min} & {Max}\\
\hline
$h_1 \to bb$ & \SI{-49.18}{\percent}&\SI{28.33}{\percent}& \SI{-48.31}{\percent}& \SI{37.35}{\percent}& \SI{-49.18}{\percent}&\SI{28.33}{\percent}& \SI{-48.31}{\percent}& \SI{37.35}{\percent}\\
$h_1\to \tau \tau $ & \SI{-52.03}{\percent}&\SI{27.77}{\percent}& \SI{-51.16}{\percent}& \SI{37.53}{\percent}& \SI{-52.03}{\percent}&\SI{27.77}{\percent}& \SI{-51.16}{\percent}& \SI{37.53}{\percent}\\
$h_1 \to AA$ &\SI{-0.1}{\percent}&\SI{7.03}{\percent}&\SI{-0.1}{\percent}&\SI{7.03}{\percent}&\SI{-3.03}{\percent}&\SI{8.11}{\percent}&\SI{-0.56}{\percent}&\SI{9.7}{\percent}\\
$h_2 \to bb$ &\SI{-10.92}{\percent}&\SI{-1.29}{\percent}&\SI{-10.52}{\percent}&\SI{-1.3}{\percent}&\SI{-10.92}{\percent}&\SI{-1.3}{\percent}&\SI{-10.52}{\percent}&\SI{-1.3}{\percent}\\
$h_2 \to \tau \tau$ &\SI{-14.12}{\percent}&\SI{-4.5}{\percent}&\SI{-13.71}{\percent}&\SI{-4.5}{\percent}&\SI{-14.12}{\percent}&\SI{-4.5}{\percent}&\SI{-13.71}{\percent}&\SI{-4.5}{\percent}\\
$h_2 \to h_1 h_1$ & \SI{-72.5}{\percent}& $>$\SI{100}{\percent}& \SI{-58.37}{\percent}& $>$\SI{100}{\percent}& \SI{-30.29}{\percent}& $>$\SI{100}{\percent}& \SI{-30.29}{\percent}& $>$\SI{100}{\percent}\\
$h_2 \to AA$ & \SI{-52.58}{\percent}&\SI{8.8}{\percent}& \SI{-47.7}{\percent}&\SI{4.97}{\percent}& \SI{-40.94}{\percent}&\SI{5.82}{\percent}& \SI{-40.94}{\percent}&\SI{5.82}{\percent}\\
\hline
\end{tabular}
\caption{Scenario II: Same as Tab.~\ref{tab:scenIall}, but for scenario II, {\it
    i.e.}~$m_{h_2}= 125.09$~GeV. 
\label{tab:scenIIall}}
\end{table}

As can be inferred from Tab.~\ref{tab:scenIall}, for $h_1 = h_{125}$
and $v_S$ renormalization through the on-shell decay processes
$h_{1,2} \to AA$ the relative NLO EW corrections are of typical size
with at most 25\%, apart from the corrections to the decay $h_2 \to
h_1 h_1$ in case the decay $h_2 \to AA$ is chosen for the
renormalization of $v_S$. The reason for this is two-fold. Large
relative corrections appear due to a very small LO width 
resulting from parameter points where $\tan \alpha \approx v_S/v$. Since the case
$h_1 = h_{125}$ requires small $\alpha$ values (below about 0.37) in order to
comply with the experimental constraints, this requires $v_s \lsim 
90 \mbox{GeV}$. At the same time relatively small $v_S$ values imply
large trilinear couplings $g_{h_1h_2h_2}$,  $g_{h_2h_2h_2}$ and $g_{h_2 AA}$ (remind
that $\cos \alpha \approx 1$ for $h_1 = h_{125}$).\footnote{In the
  more restricted sample of \cite{Egle:2022wmq}, which only considered
parameter sets where SM-like Higgs decays into DM particles are
kinematically possible we did not encounter these scenarios as they
are excluded then due to the direct DM detection
constraints.} In case we use $h_1 \to AA$ for the
renormalization we are kinematically more constrained and also have to
comply with the DM constraints so that points with vanishing LO coupling can no longer occur in the sample and we do not have large corrections here. \s


When we look at the corrections for scenario I using ZEM
renormalization of $v_S$, given in Tab.~\ref{tab:scenIall}, we see
that the relative corrections to the $h_2$ decays apart from $h_2 \to
AA$ in the ZEMproc2 scheme become large. The reasons 
can be as above small LO widths combined with large
couplings. Additionally,  threshold effects in the derivative of the
$B_0$ function \cite{Passarino:1978jh} can appear at $2 m_A - m_{h_2} \gsim
0$. Such kinematic scenarios can only appear in the ZEM scheme where
for the renormalization, we are not restricted any more to 
scenarios with on-shell $h_{1,2} \to AA$ decays.  These threshold
effects increase the diagonal wave function
renormalization constants leading to large corrections. In the decays
$h_2 \to AA$ we cannot have these threshold effects, as $h_2$ must
decay on-shell into $AA$. 
Comparing the correction to $h_2 \to AA$ in the 'proc1' and 'proc2'
renormalization scheme, we see that in the latter they are of moderate
size, whereas for 'proc1' renormalization they become large. The
reason is that the counterterm $\delta v_S$ in the ZEM scheme can become large
when there is a large mass difference between the initial and final
state particles in the process used for renormalization in the ZEM
scheme. \s

Note that in  the ZEM scheme for all but the Higgs-to-Higgs decays the
sizes of the relative 
corrections are the same independently of the used process for the
renormalization of $v_S$ since these decays do not necessitate the
renormalization of $v_S$. In the OS case for the renormalization of
$v_S$, {\it cf.}~Tab.\ref{tab:scenIall}, they differ, however,
because due to the kinematic restriction 
we have different parameter samples depending on which process we use
for the renormalization of $v_S$. \s

In scenario II, where $h_2 = h_{125}$, we infer from
Tab.~\ref{tab:scenIIall} (upper) that now also the decay $h_2 \to h_1 h_1$
can get large relative corrections in case $h_1 \to AA$ is used for
renormalization. The reason for this enhancement is a vanishing LO
decay width for $\tan \alpha = v_S/v$. Since in scenario II $\alpha
\approx \pi/2$, this does not require $v_S$ to be small and hence such
parameter scenarios are not in conflict with the DM constraints. 
Note finally, that the $h_2$ decays
into $WW,$ $ZZ$ and $tt$ are kinematically closed in scenario II, and
we do not consider off-shell decays when we compute the EW corrections. \s

If we use $h_2 \to AA$ for the renormalization of $v_S$ 
then the $h_1 \to bb$ and $\tau \tau$ decays
can get large relative corrections. This can be due to threshold
effects for $2 m_A - m_{h_1} \gsim 0$ or due to large couplings
between scalars.
 For the $h_1 \to AA$ decay, the kinematics required for this
  process combined with  the experimental constraints leads to no
  strong enhancement of the relative corrections. 
The reasons for large relative
corrections to the $h_2 \to h_1 h_1$ decays are a vanishing LO width
or large couplings (which in contrast to scenario I are not
correlated with a vanishing LO width any more) or threshold effects
for $2 m_A - m_{h_1} \gsim 0$. \s

If we choose ZEM renormalization, {\it cf.}~Tab.~\ref{tab:scenIIall}, 
then $h_1 \to bb$ and $h_1 \to \tau 
\tau$ get large corrections mainly due to threshold effects in the
derivative of the B0 function for $2 m_A - m_{h_1} \gsim 0$. This is
independent of the chosen process for the $v_S$ renormalization as it does
not enter in these processes. We also have a parameter point here where the large
corrections are due to large couplings between scalar particles.
The $h_2$ decays only show large corrections in the decays into scalar
particles. For the $h_2 \to h_1 h_1$ decays the reasons can be a small
LO width, large couplings or threshold effects. 
In the case of the $h_2 \to AA$ decays we cannot have threshold
effects from $2m_A - m_{h_2} \gsim 0$ as they are kinematically not
possible. If we use $h_1 \to AA$ for the renormalization of $v_S$  
threshold effects for $2m_A - m_{h_1} \gsim 0$ can appear and lead to large
corrections. Large couplings between scalar particles can additionally
enhance the corrections independently of the used renormalization
process for $v_S$. 

\subsection{Size of NLO EW Corrections after Cuts}
To confirm the above observations we excluded in a next step from our parameter
sample all parameter points where $0\mbox{ GeV} \le 2m_A-m_{h_i}  \le 9$~GeV
($i=1,2$) to avoid large threshold corrections, and points with
$\lambda_{ijkl} > 4\pi$, where $\lambda_{ijkl}$ stands for all
possible quartic couplings between the 
scalars of our model. This eliminates possibly large couplings (which
were not eliminated yet by the constraints from perturbative
unitarity). The resulting relative NLO corrections are summarized in
Tab.~\ref{tab:aftercutsscenI} for scenario I and in
Tab.~\ref{tab:aftercutsscenII} for scenario II. \s

\begin{table}[h!]
\centering
\begin{tabular}{|c||c|c||c|c||c|c||c|c|}
\hline
& \multicolumn{2}{c||}{OSproc1-OS}& \multicolumn{2}{c||}{OSproc1-$p_*$}& \multicolumn{2}{c||}{OSproc2-OS}& \multicolumn{2}{c|}{OSproc2-$p_*$}\\
 & {Min} & {Max} & {Min} & {Max} & {Min} & {Max} & {Min} & {Max}\\
\hline
$h_1 \to bb$ &\SI{-1.45}{\percent}&\SI{0.0}{\percent}&\SI{-1.41}{\percent}&\SI{0.0}{\percent}&\SI{-1.56}{\percent}&\SI{0.0}{\percent}&\SI{-1.47}{\percent}&\SI{0.0}{\percent}\\
$h_1\to \tau \tau $ &\SI{-4.65}{\percent}&\SI{0.0}{\percent}&\SI{-4.61}{\percent}&\SI{0.0}{\percent}&\SI{-4.74}{\percent}&\SI{0.0}{\percent}&\SI{-4.66}{\percent}&\SI{0.0}{\percent}\\
$h_1 \to AA$ &\SI{-0.11}{\percent}&\SI{0.11}{\percent}&\SI{-0.11}{\percent}&\SI{0.11}{\percent}&\SI{-2.12}{\percent}&\SI{6.27}{\percent}&\SI{-2.24}{\percent}&\SI{5.88}{\percent}\\
$h_2 \to bb$ &\SI{-9.19}{\percent}&\SI{3.96}{\percent}&\SI{-9.57}{\percent}&\SI{4.01}{\percent}&\SI{-10.18}{\percent}&\SI{6.6}{\percent}&\SI{-10.52}{\percent}&\SI{8.57}{\percent}\\
$h_2 \to \tau \tau$ &\SI{-16.3}{\percent}&\SI{0.0}{\percent}&\SI{-16.79}{\percent}&\SI{0.0}{\percent}&\SI{-17.09}{\percent}&\SI{2.47}{\percent}&\SI{-17.7}{\percent}&\SI{4.97}{\percent}\\
$h_2 \to tt$ &\SI{-6.1}{\percent}&\SI{5.35}{\percent}&\SI{-6.17}{\percent}&\SI{5.21}{\percent}&\SI{-6.1}{\percent}&\SI{7.26}{\percent}&\SI{-6.16}{\percent}&\SI{6.89}{\percent}\\
$h_2 \to WW$ &\SI{-9.76}{\percent}&\SI{9.31}{\percent}&\SI{-9.85}{\percent}&\SI{9.5}{\percent}&\SI{-9.8}{\percent}&\SI{10.33}{\percent}&\SI{-9.89}{\percent}&\SI{10.82}{\percent}\\
$h_2 \to ZZ$ &\SI{-3.46}{\percent}&\SI{4.44}{\percent}&\SI{-3.51}{\percent}&\SI{4.32}{\percent}&\SI{-3.49}{\percent}&\SI{7.38}{\percent}&\SI{-3.59}{\percent}&\SI{7.73}{\percent}\\
$h_2 \to h_1 h_1$ &\SI{-6.35}{\percent}&\SI{6.14}{\percent}&\SI{-6.71}{\percent}&\SI{5.43}{\percent}& \SI{-54.97}{\percent}&\SI{10.89}{\percent}& \SI{-54.2}{\percent}&\SI{10.51}{\percent}\\
$h_2 \to AA$ &\SI{-6.25}{\percent}&\SI{2.14}{\percent}&\SI{-5.9}{\percent}&\SI{2.25}{\percent}&\SI{-0.14}{\percent}&\SI{0.13}{\percent}&\SI{-0.12}{\percent}&\SI{0.13}{\percent}\\
\hline
\hline
& \multicolumn{2}{c||}{ZEMproc1-OS}& \multicolumn{2}{c||}{ZEMproc1-$p_*$}& \multicolumn{2}{c||}{ZEMproc2-OS}& \multicolumn{2}{c|}{ZEMproc2-$p_*$}\\
 & {Min} & {Max} & {Min} & {Max} & {Min} & {Max} & {Min} & {Max}\\
\hline
$h_1 \to bb$ &\SI{-1.56}{\percent}&\SI{-1.1}{\percent}&\SI{-1.47}{\percent}&\SI{-1.04}{\percent}&\SI{-1.56}{\percent}&\SI{-1.1}{\percent}&\SI{-1.47}{\percent}&\SI{-1.04}{\percent}\\
$h_1\to \tau \tau $ &\SI{-4.74}{\percent}&\SI{-4.29}{\percent}&\SI{-4.66}{\percent}&\SI{-4.23}{\percent}&\SI{-4.74}{\percent}&\SI{-4.29}{\percent}&\SI{-4.66}{\percent}&\SI{-4.23}{\percent}\\
$h_1 \to AA$ &\SI{-2.46}{\percent}&\SI{4.54}{\percent}&\SI{-2.46}{\percent}&\SI{4.49}{\percent}&\SI{-4.63}{\percent}&\SI{4.72}{\percent}&\SI{-5.0}{\percent}&\SI{4.58}{\percent}\\
$h_2 \to bb$ &\SI{-12.94}{\percent}&\SI{6.62}{\percent}&\SI{-13.36}{\percent}&\SI{8.54}{\percent}&\SI{-12.94}{\percent}&\SI{6.6}{\percent}&\SI{-13.36}{\percent}&\SI{8.55}{\percent}\\
$h_2 \to \tau \tau$ &\SI{-28.83}{\percent}&\SI{2.48}{\percent}&\SI{-29.2}{\percent}&\SI{4.97}{\percent}&\SI{-28.8}{\percent}&\SI{2.46}{\percent}&\SI{-29.21}{\percent}&\SI{4.98}{\percent}\\
$h_2 \to tt$ &\SI{-18.95}{\percent}&\SI{7.25}{\percent}&\SI{-18.94}{\percent}&\SI{6.89}{\percent}&\SI{-18.95}{\percent}&\SI{7.24}{\percent}&\SI{-18.94}{\percent}&\SI{6.86}{\percent}\\
$h_2 \to WW$ &\SI{-22.57}{\percent}&\SI{10.31}{\percent}&\SI{-22.56}{\percent}&\SI{10.81}{\percent}&\SI{-22.58}{\percent}&\SI{10.31}{\percent}&\SI{-22.57}{\percent}&\SI{10.8}{\percent}\\
$h_2 \to ZZ$ &\SI{-16.27}{\percent}&\SI{7.41}{\percent}&\SI{-16.26}{\percent}&\SI{7.65}{\percent}&\SI{-16.26}{\percent}&\SI{7.39}{\percent}&\SI{-16.25}{\percent}&\SI{7.69}{\percent}\\
$h_2 \to h_1 h_1$ & \SI{-49.81}{\percent}& \SI{89.19}{\percent}& \SI{-50.0}{\percent}& \SI{88.83}{\percent}& \SI{-55.46}{\percent}&\SI{11.58}{\percent}& \SI{-54.68}{\percent}&\SI{11.58}{\percent}\\
$h_2 \to AA$ & \SI{-99.98}{\percent}&\SI{5.65}{\percent}& \SI{-99.99}{\percent}&\SI{5.82}{\percent}&\SI{-15.22}{\percent}&\SI{6.13}{\percent}&\SI{-15.23}{\percent}&\SI{6.11}{\percent}\\
\hline
\end{tabular}
\caption{Scenario I after cuts: Relative NLO corrections $\delta_{\text{EW}} (h_i \to XX)$
  for all possible decay processes in 
  scenario I, {\it i.e.}~$m_{h_1}= 125.09$~GeV, and different renormalization
  schemes $\mbox{ren}_{v_S}$-$\mbox{ren}_\alpha$ after applying the
  cuts described in the text. \label{tab:aftercutsscenI}}
\end{table}

As can be inferred from Tab.~\ref{tab:aftercutsscenI}, in scenario I now all
relative corrections are relatively small, at most 29\%. Only the
relative corrections to the $h_2 \to h_1 h_1$ width 
in the OSproc2 and the ZEM schemes and the $h_2 \to AA$ decays in the
ZEMproc1 scheme can become large. The large relative corrections for
$h_2 \to h_1 h_1$ are due to a small LO width that also entails large
couplings between the 
scalar particles (see discussion above). In the OSproc1 scheme the
relative corrections are
small also for this decay channel as here the additional kinematic
constraint $2 m_A < m_{h_1} = 125.09$~GeV allows for DM decays so that
additional experimental constraints have to be considered such that
the conditions for the large corrections are not met. Additionally, in
both decay channels we can have large corrections 
in the ZEMproc1 scheme because of the counterterm $\delta v_S$ which
can become large when there is a large mass difference between the initial and final
state particles in the process $h_1 \to AA$ used for its renormalization. \s

In scenario II, {\it cf.}~Tab.~\ref{tab:aftercutsscenII}, all relative
corrections remain below 19\% apart from 
those to the $h_2 \to h_1 h_1$ decays. They are due to small LO
widths. Additionally, in the ZEMproc1 and ZEMproc2 scheme the NLO
corrections can become large again due to the counterterm $\delta v_S$
that can become large when there is a large mass difference between the initial and final
state particles in the process used for its renormalization (here $h_1
\to AA$ or $h_2 \to AA$). \s

\begin{table}
\centering
\begin{tabular}{|c||c|c||c|c||c|c||c|c|}
\hline
& \multicolumn{2}{c||}{OSproc1-OS}& \multicolumn{2}{c||}{OSproc1-$p_*$}& \multicolumn{2}{c||}{OSproc2-OS}& \multicolumn{2}{c|}{OSproc2-$p_*$}\\
 & {Min} & {Max} & {Min} & {Max} & {Min} & {Max} & {Min} & {Max}\\
\hline
$h_1 \to bb$ &\SI{-1.37}{\percent}&\SI{4.46}{\percent}&\SI{-1.38}{\percent}&\SI{0.81}{\percent}&\SI{-2.81}{\percent}&\SI{11.99}{\percent}&\SI{-1.36}{\percent}&\SI{18.5}{\percent}\\
$h_1\to \tau \tau $ &\SI{-4.47}{\percent}&\SI{3.96}{\percent}&\SI{-4.47}{\percent}&\SI{1.68}{\percent}&\SI{-4.47}{\percent}&\SI{11.46}{\percent}&\SI{-4.5}{\percent}&\SI{18.96}{\percent}\\
$h_1 \to AA$ &\SI{-0.1}{\percent}&\SI{0.08}{\percent}&\SI{-0.1}{\percent}&\SI{0.09}{\percent}&\SI{-4.68}{\percent}&\SI{1.01}{\percent}&\SI{-1.24}{\percent}&\SI{1.53}{\percent}\\
$h_2 \to bb$ &\SI{-1.39}{\percent}&\SI{0.0}{\percent}&\SI{-1.39}{\percent}&\SI{0.0}{\percent}&\SI{-1.4}{\percent}&\SI{0.0}{\percent}&\SI{-1.4}{\percent}&\SI{0.0}{\percent}\\
$h_2 \to \tau \tau$ &\SI{-4.57}{\percent}&\SI{0.0}{\percent}&\SI{-4.59}{\percent}&\SI{0.0}{\percent}&\SI{-4.6}{\percent}&\SI{0.0}{\percent}&\SI{-4.59}{\percent}&\SI{0.0}{\percent}\\
$h_2 \to h_1 h_1$ &\SI{-10.6}{\percent}& \SI{46.49}{\percent}&\SI{-10.76}{\percent}& \SI{46.55}{\percent}&\SI{-11.64}{\percent}& $>$\SI{100}{\percent}&\SI{-11.7}{\percent}& $>$\SI{100}{\percent}\\
$h_2 \to AA$ &\SI{-1.04}{\percent}&\SI{4.67}{\percent}&\SI{-1.53}{\percent}&\SI{1.22}{\percent}&\SI{-0.11}{\percent}&\SI{0.11}{\percent}&\SI{-0.11}{\percent}&\SI{0.11}{\percent}\\
\hline
\hline
& \multicolumn{2}{c||}{ZEMproc1-OS}& \multicolumn{2}{c||}{ZEMproc1-$p_*$}& \multicolumn{2}{c||}{ZEMproc2-OS}& \multicolumn{2}{c|}{ZEMproc2-$p_*$}\\
 & {Min} & {Max} & {Min} & {Max} & {Min} & {Max} & {Min} & {Max}\\
\hline
$h_1 \to bb$ &\SI{-2.81}{\percent}&\SI{11.99}{\percent}&\SI{-1.36}{\percent}&\SI{18.5}{\percent}&\SI{-2.81}{\percent}&\SI{11.99}{\percent}&\SI{-1.38}{\percent}&\SI{18.5}{\percent}\\
$h_1\to \tau \tau $ &\SI{-4.46}{\percent}&\SI{11.46}{\percent}&\SI{-4.48}{\percent}&\SI{18.96}{\percent}&\SI{-4.47}{\percent}&\SI{11.46}{\percent}&\SI{-4.48}{\percent}&\SI{18.96}{\percent}\\
$h_1 \to AA$ &\SI{-0.1}{\percent}&\SI{2.54}{\percent}&\SI{-0.1}{\percent}&\SI{2.54}{\percent}&\SI{-3.03}{\percent}&\SI{2.62}{\percent}&\SI{-0.56}{\percent}&\SI{3.16}{\percent}\\
$h_2 \to bb$ &\SI{-5.6}{\percent}&\SI{-1.29}{\percent}&\SI{-5.48}{\percent}&\SI{-1.3}{\percent}&\SI{-5.6}{\percent}&\SI{-1.31}{\percent}&\SI{-5.48}{\percent}&\SI{-1.3}{\percent}\\
$h_2 \to \tau \tau$ &\SI{-8.79}{\percent}&\SI{-4.5}{\percent}&\SI{-8.69}{\percent}&\SI{-4.5}{\percent}&\SI{-8.79}{\percent}&\SI{-4.5}{\percent}&\SI{-8.69}{\percent}&\SI{-4.5}{\percent}\\
$h_2 \to h_1 h_1$ & \SI{-72.5}{\percent}& $>$\SI{100}{\percent}& \SI{-58.37}{\percent}& $>$\SI{100}{\percent}& \SI{-30.29}{\percent}& $>$\SI{100}{\percent}& \SI{-30.29}{\percent}& $>$\SI{100}{\percent}\\
$h_2 \to AA$ &\SI{-17.38}{\percent}&\SI{5.77}{\percent}&\SI{-12.06}{\percent}&\SI{4.02}{\percent}&\SI{-13.54}{\percent}&\SI{2.83}{\percent}&\SI{-13.54}{\percent}&\SI{2.83}{\percent}\\
\hline
\end{tabular}
\caption{Scenario II after cuts: Relative NLO corrections $\delta_{\text{EW}} (h_i \to XX)$
  for all possible decay processes in 
  scenario II, {\it i.e.}~$m_{h_2}= 125.09$~GeV, and different renormalization
  schemes $\mbox{ren}_{v_S}$-$\mbox{ren}_\alpha$ after applying the
  cuts described in the text. \label{tab:aftercutsscenII}}
\end{table}

We can summarize: For $h_1 = h_{125}$, the relative corrections to both $h_1$ and $h_2$
decays are in general decent being at most 20 to 30\% as long as OS
conditions are applied in the process-dependent renormalization of
$v_S$. The exception are the relative corrections to the decay $h_2
\to h_1 h_1$ which can become large due to small LO width entailing
also large couplings among the scalars. For $h_2 = h_{125}$ also the
corrections to $h_1$ decays into $bb$ and $\tau\tau$ can get large due
to threshold effects or large couplings between the scalars. If we use
the ZEM scheme for the process-dependent renormalization of $v_S$, in
scenario I all $h_2$ decays get large corrections, in scenario II the
$h_1$ decays into $bb$ and $\tau\tau$ and the $h_2$ decays into scalar
pairs get large corrections. For a better perturbative convergence, it
is hence advisable to use the OS scheme in the process-dependent
renormalization of $v_S$. However, this also restricts the possible
parameter scenarios that can be used, as the kinematic constraints for
the OS decays used for renormalization have to be met. We therefore
use the ZEM scheme, more specifically the ZEMproc2 scheme, in the
following as our standard scheme for the NLO corrections.


\subsection{Theoretical Uncertainty}
We can get an estimate of the theoretical uncertainty due to missing
higher-order corrections by investigating the NLO results in different
renormalization schemes. For this comparison to be meaningful, we have
to consistently convert the input parameters of scheme $a$ to scheme
$b$ when moving from scheme $a$ to scheme $b$. For definiteness, in
this investigation our starting scheme $a$ is ZEMproc2-OS with the
input parameters assumed to be given in this scheme. We then convert
the input parameters to 
the other schemes under investigation and compute the higher-order
corrections in these schemes. For the conversion of the input
parameters we use an approximate formula based on the bare parameter
$p_0$ which was given in Eq.~(\ref{eq:schemechange}). \s

\begin{table}[tb]
\centering
\begin{tabular}{|c||c|c||c|c||c|c||c|c|}
\hline
& \multicolumn{2}{c||}{OSproc1-OS}& \multicolumn{2}{c||}{OSproc1-$p_*$}& \multicolumn{2}{c||}{OSproc2-OS}& \multicolumn{2}{c|}{OSproc2-$p_*$}\\
\hline
$\alpha$ &\multicolumn{2}{c||}{\SI{0.1654}{}}&\multicolumn{2}{c||}{\SI{0.1655}{}}&\multicolumn{2}{c||}{\SI{0.1654}{}}&\multicolumn{2}{c|}{\SI{0.1655}{}}\\
$v_S$ &\multicolumn{2}{c||}{\SI{439.21}{\giga \electronvolt}}&\multicolumn{2}{c||}{\SI{439.37}{\giga \electronvolt}}&\multicolumn{2}{c||}{\SI{444.64}{\giga \electronvolt}}&\multicolumn{2}{c|}{\SI{444.64}{\giga \electronvolt}}\\
\hline
& $\delta_{\text{EW}}$  & $\delta_{\text{ren}}$& $\delta_{\text{EW}}$  & $\delta_{\text{ren}}$& $\delta_{\text{EW}}$  & $\delta_{\text{ren}}$& $\delta_{\text{EW}}$  & $\delta_{\text{ren}}$\\
\hline
$h_1 \to bb$ &\SI{-1.37}{\percent}&\SI{0.0}{\percent}&\SI{-1.36}{\percent}&\SI{0.0}{\percent}&\SI{-1.37}{\percent}&\SI{0.0}{\percent}&\SI{-1.36}{\percent}&\SI{0.0}{\percent}\\
$h_1\to \tau \tau $ &\SI{-4.56}{\percent}&\SI{0.0}{\percent}&\SI{-4.56}{\percent}&\SI{0.0}{\percent}&\SI{-4.56}{\percent}&\SI{0.0}{\percent}&\SI{-4.56}{\percent}&\SI{0.0}{\percent}\\
$h_1 \to AA$ &\SI{-0.0}{\percent}&\SI{-3.01}{\percent}&\SI{-0.0}{\percent}&\SI{-3.01}{\percent}&\SI{2.44}{\percent}&\SI{-0.64}{\percent}&\SI{2.37}{\percent}&\SI{-0.71}{\percent}\\
$h_2 \to bb$ &\SI{-1.23}{\percent}&\SI{-0.0}{\percent}&\SI{-1.3}{\percent}&\SI{-0.07}{\percent}&\SI{-1.23}{\percent}&\SI{-0.0}{\percent}&\SI{-1.3}{\percent}&\SI{-0.07}{\percent}\\
$h_2 \to \tau \tau$ &\SI{-14.16}{\percent}&\SI{-0.0}{\percent}&\SI{-14.23}{\percent}&\SI{-0.08}{\percent}&\SI{-14.16}{\percent}&\SI{-0.0}{\percent}&\SI{-14.23}{\percent}&\SI{-0.08}{\percent}\\
$h_2 \to tt$ &\SI{-2.23}{\percent}&\SI{0.01}{\percent}&\SI{-2.3}{\percent}&\SI{-0.06}{\percent}&\SI{-2.24}{\percent}&\SI{0.0}{\percent}&\SI{-2.31}{\percent}&\SI{-0.07}{\percent}\\
$h_2 \to WW$ &\SI{-1.71}{\percent}&\SI{0.01}{\percent}&\SI{-1.78}{\percent}&\SI{-0.06}{\percent}&\SI{-1.72}{\percent}&\SI{0.0}{\percent}&\SI{-1.79}{\percent}&\SI{-0.07}{\percent}\\
$h_2 \to ZZ$ &\SI{1.46}{\percent}&\SI{0.01}{\percent}&\SI{1.39}{\percent}&\SI{-0.06}{\percent}&\SI{1.46}{\percent}&\SI{0.0}{\percent}&\SI{1.39}{\percent}&\SI{-0.07}{\percent}\\
$h_2 \to h_1 h_1$ &\SI{0.72}{\percent}&\SI{0.25}{\percent}&\SI{0.66}{\percent}&\SI{0.19}{\percent}&\SI{0.52}{\percent}&\SI{0.05}{\percent}&\SI{0.46}{\percent}&\SI{-0.01}{\percent}\\
$h_2 \to AA$ &\SI{-2.46}{\percent}&\SI{-3.11}{\percent}&\SI{-2.39}{\percent}&\SI{-3.04}{\percent}&\SI{0.0}{\percent}&\SI{-0.67}{\percent}&\SI{0.0}{\percent}&\SI{-0.67}{\percent}\\
\hline
\hline
& \multicolumn{2}{c||}{ZEMproc1-OS}& \multicolumn{2}{c||}{ZEMproc1-$p_*$}& \multicolumn{2}{c||}{ZEMproc2-OS}& \multicolumn{2}{c|}{ZEMproc2-$p_*$}\\
\hline
$\alpha$ &\multicolumn{2}{c||}{\SI{0.1654}{}}&\multicolumn{2}{c||}{\SI{0.1655}{}}&\multicolumn{2}{c||}{\SI{0.1654}{}}&\multicolumn{2}{c|}{\SI{0.1655}{}}\\
$v_S$ &\multicolumn{2}{c||}{\SI{447.05}{\giga \electronvolt}}&\multicolumn{2}{c||}{\SI{447.21}{\giga \electronvolt}}&\multicolumn{2}{c||}{\SI{446.14}{\giga \electronvolt}}&\multicolumn{2}{c|}{\SI{446.14}{\giga \electronvolt}}\\
\hline
& $\delta_{\text{EW}}$  & $\delta_{\text{ren}}$& $\delta_{\text{EW}}$  & $\delta_{\text{ren}}$& $\delta_{\text{EW}}$  & $\delta_{\text{ren}}$& $\delta_{\text{EW}}$  & $\delta_{\text{ren}}$\\
\hline
$h_1 \to bb$ &\SI{-1.37}{\percent}&\SI{0.0}{\percent}&\SI{-1.36}{\percent}&\SI{0.0}{\percent}&\SI{-1.37}{\percent}&\SI{0.0}{\percent}&\SI{-1.36}{\percent}&\SI{0.0}{\percent}\\
$h_1\to \tau \tau $ &\SI{-4.56}{\percent}&\SI{-0.0}{\percent}&\SI{-4.56}{\percent}&\SI{0.0}{\percent}&\SI{-4.56}{\percent}&\SI{0.0}{\percent}&\SI{-4.56}{\percent}&\SI{0.0}{\percent}\\
$h_1 \to AA$ &\SI{3.51}{\percent}&\SI{0.4}{\percent}&\SI{3.51}{\percent}&\SI{0.4}{\percent}&\SI{3.11}{\percent}&\SI{0.0}{\percent}&\SI{3.03}{\percent}&\SI{-0.07}{\percent}\\
$h_2 \to bb$ &\SI{-1.23}{\percent}&\SI{0.0}{\percent}&\SI{-1.3}{\percent}&\SI{-0.07}{\percent}&\SI{-1.23}{\percent}&\SI{0.0}{\percent}&\SI{-1.3}{\percent}&\SI{-0.07}{\percent}\\
$h_2 \to \tau \tau$ &\SI{-14.15}{\percent}&\SI{0.0}{\percent}&\SI{-14.22}{\percent}&\SI{-0.08}{\percent}&\SI{-14.15}{\percent}&\SI{0.0}{\percent}&\SI{-14.22}{\percent}&\SI{-0.08}{\percent}\\
$h_2 \to tt$ &\SI{-2.25}{\percent}&\SI{-0.0}{\percent}&\SI{-2.32}{\percent}&\SI{-0.07}{\percent}&\SI{-2.25}{\percent}&\SI{0.0}{\percent}&\SI{-2.32}{\percent}&\SI{-0.07}{\percent}\\
$h_2 \to WW$ &\SI{-1.72}{\percent}&\SI{-0.0}{\percent}&\SI{-1.79}{\percent}&\SI{-0.07}{\percent}&\SI{-1.72}{\percent}&\SI{0.0}{\percent}&\SI{-1.79}{\percent}&\SI{-0.07}{\percent}\\
$h_2 \to ZZ$ &\SI{1.45}{\percent}&\SI{-0.0}{\percent}&\SI{1.38}{\percent}&\SI{-0.07}{\percent}&\SI{1.45}{\percent}&\SI{0.0}{\percent}&\SI{1.38}{\percent}&\SI{-0.07}{\percent}\\
$h_2 \to h_1 h_1$ &\SI{0.43}{\percent}&\SI{-0.03}{\percent}&\SI{0.37}{\percent}&\SI{-0.1}{\percent}&\SI{0.47}{\percent}&\SI{0.0}{\percent}&\SI{0.41}{\percent}&\SI{-0.06}{\percent}\\
$h_2 \to AA$ &\SI{1.08}{\percent}&\SI{0.41}{\percent}&\SI{1.16}{\percent}&\SI{0.48}{\percent}&\SI{0.67}{\percent}&\SI{0.0}{\percent}&\SI{0.67}{\percent}&\SI{0.0}{\percent}\\
\hline
\end{tabular}
\caption{{\tt BP1:} Relative corrections $\delta_{\text{EW}}$ and relative
  uncertainties $\delta_{\text{ren}}$ due to the renormalization scheme
  change at NLO EW for the various decay widths of benchmark point
  {\tt BP1} and for all applied renormalization schemes. The input scheme is
  ZEMproc2-OS. Also given are the input parameters $\alpha$ and $v_S$,
  which change with the renormalization scheme. \label{tab:uncert1} }
\end{table}

In the following, we give the results of the EW corrected decay widths
for two sample benchmark points. We define the relative uncertainty 
on the investigated Higgs decay width $\Gamma$ due to missing
higher-order corrections, estimated based on the  
renormalization scheme choice as ($b$=OSproc1-OS,OSproc1-$p_*$, OSproc2-OS,
OSproc2-$p_*$, ZEMproc1-OS, ZEMproc1-$p_*$, ZEMproc2-$p_*$) 
\beq
\delta_{\text{ren}} =
\frac{\Gamma^b-\Gamma^{\text{ZEMproc2-OS}}}{\Gamma^{\text{ZEMproc2-OS}}}
\;.
\label{eq:relrenorm}
\eeq
The first benchmark point is chosen from the scenario I sample, and is
defined by the following input parameters ($m_{h_1} = 125.09$~GeV)
\beq  
\mbox{\tt BP1: } \; m_{h_2} = 590.48  \mbox{ GeV}, \; m_A = 61.93 \mbox{ GeV} \;, v_s = 446.14 \mbox{ GeV}, \; \alpha = 0.1654 \;.  
\eeq  
Table~\ref{tab:uncert1} displays for {\tt BP1} the relative EW
corrections $\delta_{\text{EW}}$ Eq.~(\ref{eq:relewcorr}) and the relative
uncertainty $\delta_{\text{ren}}$ Eq.~(\ref{eq:relrenorm}) due to
missing higher-order corrections based on a renormalization scheme
change. The table also contains the input values which change with the
renormalization scheme, namely $v_S$ and $\alpha$. We show the results
for all applied renormalization schemes. First of all note, that
$\delta_{\text{ren}}=0$ for ZEMproc2-OS for all decays as this is our
original renormalization scheme. For all other schemes, we see that
$\alpha$ barely changes and $v_S$ changes by at most 2\% when the
renormalization scheme is altered. In line with this observation, we
see that the change of the EW corrected decay widths with the
renormalization scheme is at most 3.1\%. This is as expected for
relative EW corrections $\delta_{\text{EW}}$ that are found to be of small to moderate
size for this parameter point, not exceeding 15\%. \s

\begin{table}[tb]
\centering
\begin{tabular}{|c||c|c||c|c|}
\hline
& \multicolumn{2}{c||}{OSproc2-OS}& \multicolumn{2}{c|}{OSproc2-$p_*$}\\
\hline
$\alpha$ &\multicolumn{2}{c||}{\SI{-1.5669}{}}&\multicolumn{2}{c|}{\SI{-1.5668}{}}\\
$v_S$ &\multicolumn{2}{c||}{\SI{23.74}{\giga \electronvolt}}&\multicolumn{2}{c|}{\SI{24.49}{\giga \electronvolt}}\\
\hline
& $\Delta \Gamma$ & $\delta \Gamma$& $\Delta \Gamma$ & $\delta \Gamma$\\
\hline
$h_1 \to bb$ &\SI{13.71}{\percent}&\SI{-4.11}{\percent}&\SI{7.66}{\percent}&\SI{-9.21}{\percent}\\
$h_1\to \tau \tau $ & \SI{12.86}{\percent}&\SI{-4.14}{\percent}&\SI{6.81}{\percent}&\SI{-9.28}{\percent}\\
$h_2 \to bb$ & \SI{-1.34}{\percent}&\SI{-0.0}{\percent}&\SI{-1.34}{\percent}&\SI{-0.0}{\percent}\\
$h_2 \to \tau \tau$ & \SI{-4.54}{\percent}&\SI{0.0}{\percent}&\SI{-4.54}{\percent}&\SI{0.0}{\percent}\\
$h_2 \to AA$ & \SI{0.0}{\percent}& \SI{47.22}{\percent}&\SI{-0.0}{\percent}& \SI{47.22}{\percent}\\
\hline
\end{tabular}

\begin{tabular}{|c||c|c||c|c||c|c||c|c|}
\hline 
& \multicolumn{2}{c||}{ZEMproc1-OS}& \multicolumn{2}{c||}{ZEMproc1-$p_*$}& \multicolumn{2}{c||}{ZEMproc2-OS}& \multicolumn{2}{c|}{ZEMproc2-$p_*$}\\
\hline
$\alpha$ &\multicolumn{2}{c||}{\SI{-1.5669}{}}&\multicolumn{2}{c||}{\SI{-1.5668}{}}&\multicolumn{2}{c||}{\SI{-1.5669}{}}&\multicolumn{2}{c|}{\SI{-1.5668}{}}\\
$v_S$ &\multicolumn{2}{c||}{\SI{21.58}{\giga \electronvolt}}&\multicolumn{2}{c||}{\SI{21.58}{\giga \electronvolt}}&\multicolumn{2}{c||}{\SI{20.46}{\giga \electronvolt}}&\multicolumn{2}{c|}{\SI{21.21}{\giga \electronvolt}}\\
\hline
& $\delta_{\text{EW}}$  & $\delta_{\text{ren}}$& $\delta_{\text{EW}}$  & $\delta_{\text{ren}}$& $\delta_{\text{EW}}$  & $\delta_{\text{ren}}$& $\delta_{\text{EW}}$  & $\delta_{\text{ren}}$\\
\hline
$h_1 \to bb$ &\SI{16.66}{\percent}&\SI{-1.62}{\percent}&\SI{10.01}{\percent}&\SI{-7.24}{\percent}&\SI{18.59}{\percent}&\SI{0.0}{\percent}&\SI{10.38}{\percent}&\SI{-6.93}{\percent}\\
$h_1\to \tau \tau $ &\SI{15.81}{\percent}&\SI{-1.64}{\percent}&\SI{9.15}{\percent}&\SI{-7.29}{\percent}&\SI{17.74}{\percent}&\SI{0.0}{\percent}&\SI{9.52}{\percent}&\SI{-6.98}{\percent}\\
$h_2 \to bb$ &\SI{-1.34}{\percent}&\SI{-0.0}{\percent}&\SI{-1.34}{\percent}&\SI{-0.0}{\percent}&\SI{-1.34}{\percent}&\SI{0.0}{\percent}&\SI{-1.34}{\percent}&\SI{-0.0}{\percent}\\
$h_2 \to \tau \tau$ &\SI{-4.54}{\percent}&\SI{0.0}{\percent}&\SI{-4.54}{\percent}&\SI{0.0}{\percent}&\SI{-4.54}{\percent}&\SI{0.0}{\percent}&\SI{-4.54}{\percent}&\SI{0.0}{\percent}\\
$h_2 \to AA$ &\SI{-18.93}{\percent}&\SI{19.35}{\percent}&\SI{-25.59}{\percent}&\SI{9.55}{\percent}& \SI{-32.07}{\percent}&\SI{0.0}{\percent}&\SI{-29.83}{\percent}&\SI{3.3}{\percent}\\
\hline
\end{tabular}
\caption{{\tt BP2:} Same as Tab.~\ref{tab:uncert2}, but for {\tt BP2}. \label{tab:uncert2} }
\end{table}

We investigate a second benchmark point characterized by larger EW
corrections and theoretical uncertainties. It is chosen from the scenario II sample, and is
defined by the following input parameters ($m_{h_2} = 125.09$~GeV) 
\beq  
\mbox{\tt BP2: } \; m_{h_1} = 67.96  \mbox{ GeV}, \; m_A = 58.29 \mbox{ GeV} \;, v_s = 20.46 \mbox{ GeV}, \; \alpha = -1.5669 \;.  
\eeq  
The relative EW corrections and uncertainties in the corrected decay
widths for {\tt BP2} are shown in Table~\ref{tab:uncert2}. Since for
these parameter values an OS decay $h_1 \to AA$ is kinematically not
possible we cannot use this process for renormalization so that the EW
corrections cannot be calculated in the OSproc1-OS and OSproc1-$p_*$
schemes. In the OSproc2-OS and OSproc2-$p_*$ schemes the decay width
$h_2 \to AA$ is used for the renormalization so that the LO decay
width $\Gamma (h_2 \to AA)$ is compared to the NLO decay width,
computed in the ZEMproc2-OS scheme, for the derivation of the
theoretical uncertainty, leading 
to a relatively large value of $\delta_{\text{ren}}$ which is,
however, artificial, due to the LO-NLO comparison. We see, however,
that compared to {\tt BP1} we have larger theoretical uncertainties in
the process $h_2 \to AA$ in the ZEM schemes, ranging up to $\sim
19$\%. This can be understood by looking at the input 
parameters. The large difference between the non-SM-like Higgs mass $m_{h_1}$
and $v_S$ leads to relatively large couplings involved in the wave
function renormalization constants, increasing the counterterms, the
corrections and the relative uncertainty. The uncertainties of the
remaining NLO decays are of small to moderate size. \s

Overall, we observe that the uncertainties in the EW corrections are
small or of moderate size, apart from scenarios where large scalar
couplings are involved. Here the corrections and the related
uncertainties can become significant to large.

\subsection{Phenomenological Impact of the EW Corrections}
\subsubsection{Higgs-to-Invisible Decays}
\begin{figure}[t!]
    \centering
    \includegraphics[width=0.8\textwidth]{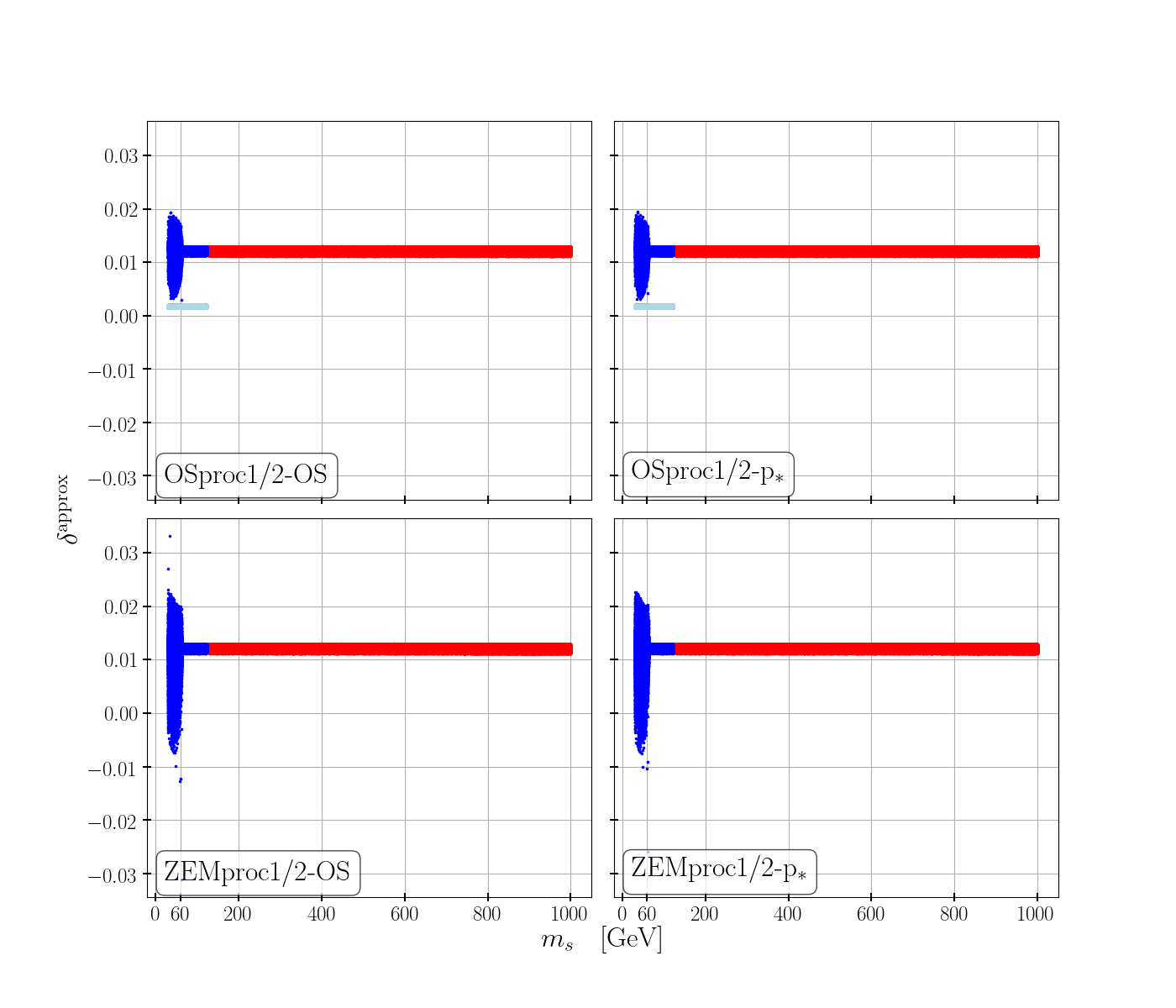}
\vspace*{-0.4cm}
    \caption{Relative uncertainty $\delta^{\text{approx}}$ in the EW NLO
      BR$(h_{125} \to AA)$ due to using the EW
      correction in the partial decay width for $h_{125} \to AA$
      only, as a function of the mass $m_S$ of the non-SM-like Higgs
      boson for scenario I (red) and II (blue points) in the four possible different
      renormalization schemes. \label{fig:approxgoodness}}
\end{figure}

We first checked if the approximate NLO branching ratio of $h_{125}$ into
a DM pair, $\mbox{BR}^{\text{NLO,approx}} (h_{125} \to AA)$,
defined in \cite{Egle:2022wmq}, where we included the NLO EW
corrections only in the decay width $\Gamma (h_{125} \to AA)$ was
sufficiently good, by comparing it with the $\mbox{BR}^{\text{NLO}}
(h_{125} \to AA) \equiv \mbox{BR} (h_{125} \to AA)$ as defined in
Eq.~(\ref{eq:bran}),  where we included the EW corrections to all
on-shell, non-loop-induced decays in the total
width. Figure~\ref{fig:approxgoodness} shows the relative difference
$\delta^{\text{approx}}$ between the two approaches, defined as
\beq
\delta^{\text{approx}} &=& \frac{\mbox{BR}^{\text{NLO}}(h_{125} \to AA) -
\mbox{BR}^{\text{NLO,approx}}(h_{125} \to
AA)}{\mbox{BR}^{\text{NLO,approx}}(h_{125} \to AA)} \;,
\eeq
as a function of the non-SM-like scalar mass $m_S$ 
for the four different
renormalization schemes OSproc1/2-OS, OSproc1/2-$p_*$, ZEMproc1/2-OS, and
ZEMproc1/2-$p_*$. The relative difference can be written as
\beq
\delta^{\text{approx}} &=& \frac{k_i^2 \Gamma^{\text{SM}}_{\text{tot}} +
  \Gamma^{\text{NLO}}_{h_{125} \to AA} + \delta \cdot
  \Gamma^{\text{LO}}_{h_{125} \to h_1h_1} }{\Gamma^{\text{NLO}}_{\text{tot}} } - 1 \;,
\eeq
where $\delta =1 \, (0)$ for $m_{h_{125}} \gsim 2 m_S \, (\lsim 2
m_s)$ and $k_i$ has been defined in
Eq.~(\ref{eq:HiggsVbosoncoupling}). Note that in the analysis of
\cite{Egle:2022wmq}, we used a sample of parameter points, where $h_{125}$
decays into DM particles are 
kinematically possible, hence $m_A \lsim m_{h_{125}}/2$. This means
that for scenario I (red points in the figure) where $h_1 \equiv
h_{125}$, the non-SM-like decay 
$h_2 \to AA$ is always possible and can be used for the renormalization of
$v_S$. In the other case, $h_2 \equiv h_{125}$, not all parameter
points necessarily fulfill the condition 
where $h_1 \to AA$ can be used for renormalization. Then no EW
corrections are calculated and both our new and the 
previous results should agree so that $\delta_{\text{approx}}=0$. 
These are the light blue points in the upper left and upper right
plots of Fig.~\ref{fig:approxgoodness} which show the results for the
case that the OS condition is used for the renormalization of
$v_S$. Using ZEM, we do have this kinematic constraint so that
$\delta^{\text{approx}} \ne 0$ for all parameter points. We remark,
that for the light blue points $\delta^{\text{approx}}$ is not exactly 
zero. This is due to a small off-set in the total widths (below 3\%)
used in \cite{Egle:2022wmq} and in this paper which stems from using
an older {\tt HDECAY} version in \cite{Egle:2022wmq}. For the blue
points (scenario II) with $m_S \le m_{h_{125}}/2$, the $h_{125} \to h_1 h_1$ decay is
open and its NLO corrections that were not included in \cite{Egle:2022wmq} play a
role in $\delta^{\text{approx}}$. We see that their effect is larger if
the ZEM scheme is applied as we already learned from our investigations
in Subsec.~\ref{subsec:size}. Nevertheless, overall the deviations
remain below $\pm 4\%$. In all remaining points, the deviations
in $\delta^{\text{approx}}$ are due to the inclusion of the EW corrections
to the fermionic decays. They remain below 1.5\%. 
The plot demonstrates that for the allowed
parameter points in both scenarios I and II the approximation used in
\cite{Egle:2022wmq} is quite good and the deviation is well below
the experimental precision. The 
application of the approximation in \cite{Egle:2022wmq} for this decay
process was hence justified. \s

\subsubsection{Impact on Allowed Parameter Regions}
In order to investigate the phenomenological impact of the EW
corrections we generated two parameter samples: Sample 1 is the
parameter sample which we also used in the previous sections. It is
based on the check of the single- and di-Higgs constraints using
the branching ratios at LO in the EW corrections. Sample 2 is the
parameter sample, where we include the EW corrections in the branching
ratios as defined in Eq.~(\ref{eq:bran}). As renormalization scheme,
we use the ZEM scheme, more specifically ZEMproc2-OS, as it allows
to use the largest amount of parameter 
scenarios\footnote{In the OS scheme, we are restricted by kinematic
  constraints.}. We do not apply any mass
cuts or cuts on the couplings to suppress large NLO corrections. We
only make sure to exclude parameter 
scenarios where the EW corrections lead to unphysical negative decay
widths. \s

We then investigate if the allowed parameter regions of sample 2
change with respect to sample 1. We find that more points are
rejected when we include the NLO corrections compared to the case
where we only take LO decay widths (see also discussion
below). Unfortunately, however, the shape of the allowed parameter
regions overall did not change. The NLO corrections hence do not have
a direct impact on the exclusion of certain parameter regions of our
model yet. With increasing
experimental precision in the future, it is evident, however, that
higher-order EW corrections have to be considered. 

\subsection{Higgs Pair Production}
Higgs pair production is one of the most prominent processes
investigated at the LHC. Its measurement allows the extraction of the
trilinear Higgs self-interaction \cite{DiMicco:2019ngk} and thereby the ultimate
experimental test of the Higgs mechanism
\cite{Djouadi:1999gv,Djouadi:1999rca}. At the LHC, the dominant Higgs
pair production process is gluon fusion into Higgs pairs
\cite{DiMicco:2019ngk,Baglio:2012np,LHCHiggsCrossSectionWorkingGroup:2016ypw}
which at LO is mediated by heavy quark triangle and box diagrams
\cite{Glover:1987nx,Dicus:1987ic,Plehn:1996wb}. The experiments search
for Higgs pairs both in resonant and non-resonant Higgs pair
production. The limits on the Higgs self-coupling between
three SM-like Higgs bosons in terms of the SM trilinear Higgs
coupling are at 95\% CL $\kappa_\lambda \in [-0.4,6.3]$ given by ATLAS
\cite{ATLAS:2022jtk} and $\kappa_\lambda \in [-1.24,6.49]$ given by
CMS \cite{CMS:2022dwd}. After applying all constraints described in
Sec.~\ref{sec:scan} we find for the still allowed $\kappa_\lambda$
values in the CxSM,
\beq
\begin{array}{ll}
\mbox{Scenario I }(h_1 \equiv h_{125}): & \kappa_\lambda \in [0.55,1.10] \\
\mbox{Scenario II }(h_2 \equiv h_{125}): & \kappa_\lambda \in [0.15,1.21] 
\end{array} \;.
\eeq
So there is still some room left to deviate from the SM value, values
equal to zero are excluded, however. When compared to other models as
{\it e.g.}~those discussed in  \cite{Abouabid:2021yvw}, we see that
the trilinear self-coupling of the SM-like Higgs in the CxSM, when
$h_1 \equiv h_{125}$, is in general more constrained than in the CP-conserving
(2HDM) and CP-violating 2-Higgs-Doublet Model (C2HDM), the
next-to-2HDM (N2HDM) or the  next-to-minimal supersymmetric extension
(NMSSM) where for some of the models the coupling can still be
zero and also take negative values. In case of scenario II, the lower
limit is below the one in the (C)2HDM and the N2HDM. 
It is interesting to note, that in the case that we do not cut
out degenerate Higgs scenarios with large mixing angles where the
discovered Higgs signal is built up by two scalar resonances close to
each other, larger deviations in the trilinear couplings would be
possible, with negative or zero trilinear coupling values. After
application of the NLO EW corrections and hence using the sample 2, we
see that the Higgs self-couplings are not further constrained compared
to sample 1. \s

While the range of allowed trilinear Higgs self-couplings
is not sensitive yet to the EW corrections on the Higgs decays, the
di-Higgs production cross sections are sensitive. More specifically,
the maximum allowed di-Higgs cross section depends on whether or not
EW corrections are included. When we use sample 1 where we calculate the 
resonant di-Higgs rate by multiplying the NNLO QCD $h_2$ production cross section
obtained from {\tt SuShi} with the LO branching ratio BR($h_2 \to
h_1 h_1$) then the maximum allowed di-Higgs cross section is given by
{\tt BP3}, defined as
\beq
\begin{array}{llll}
\multicolumn{3}{l}{\mbox{\underline{{\tt BP3}: Sample 1 - max di-Higgs at LO EW}}} \\
\mbox{Input parameters:} & m_{h_1} = 125.09 \mbox{ GeV } & m_{h_2} = 260.96 \mbox{
 GeV } & m_A = 257.36 \mbox{ GeV } \\
& \alpha = -0.312 & v_S = 87.61 \mbox{ GeV.}
\end{array}
\eeq
The SM-like di-Higgs production cross section, the trilinear
($\kappa_{h_i h_j h_k}$) and
top-Yukawa couplings ($y_{t,h_{1/2}}$) normalized to the SM values are
given by
\beq
\begin{array}{lll}
\sigma_{h_1 h_1} = 585 \mbox{ fb: } & \kappa_{h_1 h_1 h_1} = 0.781 &
  y_{t,h_1} = 0.952 \\
& \kappa_{h_2 h_1 h_1} = 1.122  & h_{t,h_2} = 0.307 
\end{array}
\eeq
For completeness, we also give the di-Higgs values for NLO branching
ratios for the various renormalization schemes. The values in brackets
are the consistently converted $v_S$ and $\alpha$ values. We have in
the ZEMproc1-OS scheme $\sigma_{h_1 h_1} = 563.56$~fb (75.28 GeV,
-0.312), in the ZEMproc1-$p_\star$ scheme $\sigma_{h_1 h_1} = 563.64$~fb (75.39 GeV,
-0.312), in the ZEMproc2-OS scheme $\sigma_{h_1 h_1} = 603.94$~fb (87.61 GeV,
-0.312), and in the ZEMproc2-$p_\star$ scheme $\sigma_{h_1 h_1} = 603.78$~fb (87.60 GeV,
-0.312). For this parameter point, the maximum relative correction of
the cross section is small 
with 3\%, with the corresponding uncertainty on the NLO corrected cross section
of 7\% being larger. \s

If on the other hand, we take the parameter values of sample 2 where
the di-Higgs constraints are applied to resonant $h_1 h_1$ production
computed with the NLO branching ratio of $h_2 \to h_1 h_1$, then we
obtain, when we apply the ZEMproc2-OS scheme, the still allowed maximum
cross section from benchmark point {\tt BP4}, defined as
\beq
\begin{array}{llll}
\multicolumn{3}{l}{\mbox{\underline{{\tt BP4}: Sample 2 - max di-Higgs at
  NLO EW (ZEMproc2-OS)}}} \\
\mbox{Input parameters:} & m_{h_1} = 125.09 \mbox{ GeV } & m_{h_2} = 262.25 \mbox{
 GeV } & m_A = 337.43 \mbox{ GeV } \\
& \alpha = -0.317 & v_S = 101.03 \mbox{ GeV,}
\end{array}
\eeq
with the following di-Higgs cross section and relevant coupling
modifiers
\beq
\begin{array}{lll}
\sigma_{h_1 h_1} = 582.84 \mbox{ fb: } & \kappa_{h_1 h_1 h_1} = 0.783 &
  y_{t,h_1} = 0.950 \\
& \kappa_{h_2 h_1 h_1} = 1.081  & h_{t,h_2} = 0.312  
\end{array}
\eeq
The di-Higgs cross section changes to 577.61~fb if we multiply the {\tt SusHi}
$h_2$ production cross section with the LO branching
ratio of the $h_2 \to h_1 h_1$ decay. 
As can be inferred from these values, the maximum allowed cross
section barely varies from non-exclusion ({\tt BP3}) to inclusion ({\tt BP4}) of the EW
corrections in the branching ratio of the di-Higgs decay. We also give
the cross sections obtained for the other renormalization schemes. We have in
the ZEMproc1-OS scheme $\sigma_{h_1 h_1} = 527.86$~fb (85.59 GeV,
-0.317), in the ZEMproc1-$p_\star$ scheme $\sigma_{h_1 h_1} = 528.01$~fb (85.72 GeV,
-0.318), and in the ZEMproc2-$p_\star$ scheme $\sigma_{h_1 h_1} = 582.63$~fb (101.02 GeV,
-0.318). For this parameter point, the maximum relative correction of
the cross section is small 
with -9\%. The corresponding uncertainty on the NLO corrected cross section
is again larger with 10\%. \s

Comparing the results for {\tt BP3} and {\tt BP4}, we remark that the
NLO value in the ZEMproc2-OS scheme of {\tt BP3} exceeds the one of
{\tt BP4}, which is the reason why {\tt BP3} is excluded in sample 2
due to the resonant di-Higgs constraints. However, it should be noted
that for these points the relative corrections are small and below the theory
uncertainties. This is also reflected in the fact that the allowed trilinear
values barely change independent if we include or not the EW
corrections in the Higgs-to-Higgs decay branching ratio. 
Overall we find, when we apply the resonant di-Higgs constraints from
the experimental searches, that less points are rejected when the NLO
branching ratio BR($h_2 \to h_1 h_1$) is included instead of the LO
one. This has to be taken with caution, however, due to remaining
theory uncertainties. And we remind that we do
not apply the resonant di-Higgs constraints when the narrow-width
approximation is not justified any more. Since the total width can be
smaller or larger depending on the parameter point and the inclusion
or not of the EW corrections in the decay widths this of course also
has an impact on the excluded region. \s

\subsection{Points with Vanishing Coupling $\lambda_{h_1 h_1 h_2}$}
The discussion so far has shown that the NLO corrections do not impact
the shape of the parameter regions that are still allowed but they affect
individual parameter points. We want to discuss now the interesting
situation of parameter configurations where the trilinear coupling 
$\lambda_{h_1 h_1 h_2}$ is close to zero. Here, the impact of the NLO
corrections could become important. We found that this is indeed the
case. For some of these parameter points the NLO corrections to $h_2
\to h_1 h_1$ shifted the point above the allowed Higgs data
constraints. As an example we  look at the benchmark point {\tt BP5},
defined as
\beq
\begin{array}{llll}
\multicolumn{3}{l}{\mbox{\underline{{\tt BP5}: Scenario 2 - small
  $\lambda_{h_1 h_1 h_2}$}}} \\
\mbox{Input parameters:} & m_{h_1} = 46.11 \mbox{ GeV } & m_{h_2} = 125.09 \mbox{
 GeV } & m_A = 60.88 \mbox{ GeV } \\
& \alpha = 1.390 & v_S = 818.10 \mbox{ GeV.}
\end{array}
\eeq  
We are close to the case $\tan\alpha=v_S/v$ where the coupling
$\lambda_{h_1 h_1 h_2}$ vanishes so that the partial decay width $\Gamma(h_2 \to h_1
h_1)$ is small. The inclusion of the NLO corrections can considerably
change this. And indeed we find, that the relative correction is (in
the ZEMproc2-OS scheme)
\beq
\delta_{\text{EW}} (h_2 \to h_1 h_1) = 0.55 \;.
\eeq
This shift is so large that the parameter point is not allowed any
more when NLO corrections are included. 


\section{Conclusions} 
\label{sec:concl}
In this paper we computed the EW corrections to the decay
widths of the visible Higgs bosons of the CxSM, the SM extended by a
complex singlet field. Applying two separate $\mathbb{Z}_2$ symmetries and
requiring one VEV to be zero, the model contains one stable DM
candidate. The EW corrections complete a previous computation of our group where
the EW corrections to the Higgs decays into a DM pair were
computed, by calculating the EW corrections to the decays into
on-shell fermionic, massive gauge boson and Higgs pair final
states. We do not compute corrections to off-shell decays nor to the 
loop-induced decays into gluon or photon pairs. We apply two different
renormalization schemes for the renormalization of the mixing angle
and four different schemes for the renormalization of the singlet
VEV. The latter is renormalized 
process-dependent both on-shell and in the so-called ZEM scheme that
we introduced in a previous calculation. The renormalization of the
mixing angle is gauge-parameter independent by applying the
alternative tadpole scheme and extracting the gauge-independent
self-energies through the pinch technique. Our computation has been
implemented in the code {\tt EWsHDECAY} which has been publicly made
available. The user has the options to choose between different
renormalization schemes, including also the possibility to convert the
input parameters, which has to be applied for a meaningful interpretation
of the change of the renormalization scheme. For the numerical analysis we performed
an extensive scan in the CxSM parameter space including theoretical
and experimental constraints as well as DM constraints. We also
considered the constraints from di-Higgs searches at the LHC. We found
that our parameter points are able to saturate the DM relic
density, and that direct search limits are already sensitive to parts
of the model for parts of the DM mass region where the collider
searches in invisible SM-like Higgs decays are not sensitive due to
kinematic constraints. \s

The analysis of the impact of the relative EW corrections showed that
they can be large in case of relatively large involved couplings, due
to threshold effects or because the LO decay width is small. Here, the
NLO width can become important and exclude parameter points that
would be allowed at LO. For parameter configurations with exactly
vanishing LO, and hence also, NLO width, we included the option to
calculate the next-to-next-to-leading-order (NNLO) width. The applied formula is,
however, only approximate when moving away from the exact zero LO
value. If we exclude parameter configurations with large couplings or
large threshold effects, the relative corrections are of typical EW size of up
to 25\%. The estimated theory uncertainty due to missing higher-order
corrections based on the change of the renormalization scheme with
converted input parameters is of a few per cent. The
phenomenological impact of the computed EW corrections is that
specific parameter points that are allowed at LO would be excluded at
NLO. The overall distribution of the allowed parameter points,
however, is not yet affected by the EW corrections. With increasing
future precision in the experiments, this will change, however, and the here
computed corrections provide an important contribution to reliably
constrain the allowed parameter space of the model. \s

We also investigated the phenomenology of di-Higgs production. Our
parameter scan revealed that deviations of the SM-like trilinear Higgs
self-coupling from the SM value are still allowed. However, vanishing or
negative trilinear coupling values are not allowed any more. The allowed range
of the trilinear coupling barely changes when we include the NLO
instead of the LO branching ratio in the Higgs-to-Higgs decay in  
resonant di-Higgs production and compare with the experimental results
on resonant di-Higgs searches. For the investigated parameter points
with maximum di-Higgs cross sections at LO, respectively at NLO, we
found that the relative corrections on the di-Higgs cross sections are
small and below the theory uncertainty. This picture may change,
however, when the full EW corrections to the complete di-Higgs
production cross section are considered. 


\iflanguage{english}

\bigskip
\bigskip
\subsubsection*{Acknowledgments}
RS and JV are supported by FCT under contracts UIDB/00618/2020, UIDP/00618/2020, PTDC/FIS-PAR/31000/2017, CERN/FISPAR
/0002/2017, CERN/FIS-PAR/0014/2019. The work of FE and MM is supported by the
BMBF-Project 05H21VKCCA. 


\providecommand{\href}[2]{#2}\begingroup\raggedright\endgroup

\end{document}